\documentclass[aps,prl,amsmath,amssymb,reprint]{revtex4-2}

\usepackage{graphicx}
\usepackage{dcolumn}
\usepackage{bm}

\begin{document}

\preprint{APS/123-QED}

\title{Heat transfer augmentation by recombination reactions in turbulent \\ reacting boundary layers at elevated pressures}%

\author{Nikolaos Perakis}
\affiliation{%
 Chair of Space Propulsion, Technical University of Munich, 85748 Garching, Germany
}%
 \affiliation{
 Department of Mechanical Engineering, Stanford University, Stanford, CA 94305, USA
}

\author{Oskar Haidn}%
\affiliation{%
 Chair of Space Propulsion, Technical University of Munich, 85748 Garching, Germany
}%

\author{Matthias Ihme}
\affiliation{
 Department of Mechanical Engineering, Stanford University, Stanford, CA 94305, USA
}%

\date{\today}

\begin{abstract}
A study of a reacting boundary layer flow with heat transfer at conditions typical for configurations at elevated pressures has been performed using a set of direct numerical simulations. Effects of wall temperatures are investigated, representative for cooled walls of gas turbines and sub-scale rocket engines operating with hydrocarbon as fuels. The results show that exothermic chemical reactions induced by the low-enthalpy in the boundary layer take place predominantly in the logarthimic sub-layer. The majority of the heat release is attributed to the exothermic recombination of OH and CO to produce CO$_2$ and H$_2$O. The recombination reactions result in an increase of the wall heat loads by up to 20\% compared to the inert flow. The gas composition experiences strong deviations from the chemical equilibrium conditions. In fact, a quenching of the major species is observed within the viscous sub-layer and the transition region. Analysis of chemical time-scales shows that the location of quenched composition coincides with the region where the Damk\"ohler number decreases below unity. Within the viscous sub-layer, a secondary reaction zone is detected, involving the production of formyl and formaldehyde radicals that provide an additional source of energy release. The analysis of the reaction paths showed that reactions with zero activation energy are responsible for this change in gas composition, which also account for the initial branching of hydrocarbon fuels decomposition according to previous auto-ignition studies. The effect of the secondary recombination reactions is more prominent for the lower wall temperature case. Finally, the role of turbulent fluctuations on the species net chemical production rates is evaluated, showing a strong correlation between species and temperature fluctuations. This leads to a pronounced deviation of the mean reaction rates $\overline{\dot{\omega}(Y_k,T)}$ from the reaction rates obtained under the assumption of laminar finite rate.

\end{abstract}

\maketitle


\section{Introduction}
\label{sec:intro}

Accurate predictions of the flame-wall interaction are crucial in the design of combustion systems. In most industrial applications that operate in confined geometries including rocket thrust chambers, gas-turbine combustors and automotive engines, flame-wall interaction strongly affects fuel consumption and pollutant formation \citep{dreizler2015advanced,mann2014transient}, while temporal and spatial fluctuations induced by it can influence the thermal loads and engine lifetime. Classical turbulent combustion models \citep{peters1984laminar} do not account for the wall effects. Hence a better understanding of the flame-wall interaction is required for the development of predictive combustion models.

A major physical phenomenon that introduces substantial uncertainties on the prediction of heat transfer is the effect of exothermic recombination reactions within the reacting boundary layer. In particular for rocket applications, modeling these reactions is especially critical in hydrocarbon combustion, where the chemical time-scales are lower than in the case of hydrogen/oxygen chemistry. Efforts to model the influence of non-adiabatic effects on the gas composition, the associated heat release and expected wall heat loads have been carried out using flamelet methods \citep{ma2018nonadiabatic,breda2019generation,perakis2018development,Cecere2011,fiorina2003modelling}.
Studies aiming at quantifying the importance of the recombination reactions on heat loads augmentation have been carried out by considering methane/oxygen mixtures in rocket combustion chambers \citep{perakis2020PCI, betti2016chemical, rahn2019conjugate, perakis2020wall}. At the same time, wall model extensions have been proposed in an effort to capture the near-wall chemical kinetics, but so far are limited to hydrogen/oxygen studies, which can be safely assumed to be in chemical equilibrium even in the cold boundary layer \citep{muto2019equilibrium}.

Despite ongoing modeling efforts, the physical mechanisms controlling turbulent flame-wall interaction, recombination reactions and the related wall heat fluxes have not been completely understood. Because of the difficulties in obtaining accurate near-wall experimental measurements, Direct Numerical Simulation (DNS) represents an alternative for studying the processes taking place in the near-wal region.

As far as high-fidelity simulations of flame-wall interaction are concerned, head-on quenching configurations have often been chosen, where premixed laminar hydrocarbon flames are propagated perpendicular to the wall. One-dimensional simulations of head-on quenching are reported in the works of Westbrook et al. \citep{westbrook1981numerical}, Hocks et al. \citep{hocks1981flame}, Popp et al. \citep{popp1996heterogeneous,popp1997analysis}. These studies concur that low-activation energy recombination reactions of chemical radicals enhance the wall heat loads and require consideration for the flame-wall interaction processes. Based on these results it is also evident that detailed chemical kinetic mechanisms are required for capturing these effects. The importance of radical recombination reactions at the wall is also emphasized in simulations of H$_2$/O$_2$ flames, as in the work of Dabireau et al. \citep{dabireau2003interaction}.

Due to the large computational costs of multi-dimensional DNS of turbulent flame-wall interaction, only few investigations have been reported. Two-dimensional DNS of head-on quenching within a reactive boundary layer was performed by Poinsot et al. \citep{poinsot1993direct} while Bruneaux et al. \citep{bruneaux1996flame} studied a three-dimensional configuration of a premixed flame propagating in constant density turbulent channel flow. Alshaalan et al. \citep{alshaalan1998turbulence} investigated sidewall quenching of a three-dimensional V-shaped premixed flame with a single-step chemistry approximation. More recently Gruber et al. \citep{gruber2010turbulent} simulated the same configuration with H$_2$/air with a detailed chemical mechanism.

As far as reacting turbulent boundary layer simulations are concerned, an even smaller number of studies can be found in literature, especially for configurations that are relevant to rocket thrust chambers. Martin et al. \citep{martin2000dns,martin2001temperature} performed DNS of hypersonic boundary layers with a single-step reaction scheme examining the feedback mechanisms between chemistry and turbulence, showing an increase in temperature fluctuations, induced by exothermic chemical reactions. Cabrit et al. \citep{cabrit2009direct} performed DNS and wall-resolved LES of multi-component mixtures in rectangular isothermal channels. Configurations with large temperature gradients and small Mach numbers were simulated, resembling rocket-like applications. However, the gas mixture that was introduced consisted to a large degree of inert nitrogen and a simplified chemical model was utilized, rendering the simulations less applicable to the thermochemical states experiences in modern methane/oxygen rocket engine hardware.

In the case of combustion devices at elevated pressures such as gas turbines and rocket combustion chambers, it is recognized that the effect of the reacting boundary layer can have a leading order effect on the performance and wall heat transfer. However, there is still insufficient understanding of the processes controlling the evolution of chemical reactions within the boundary layer, which leads to the absence of suitable models able to describe them. The present work addresses this knowledge gap in the context of recombinations in low-enthalpy environments using DNS of a reacting turbulent boundary layer. A canonical configuration is chosen for the study, with the selected operating conditions being representative for rocket combustion chamber applications. A reacting methane/oxygen gas mixture is chosen due to the relevance of methane as a fuel in the design of future space transportation vehicles.

\section{Computational setup}
\label{sec:governingEqs}

\begin{figure}
\includegraphics[width=.48\textwidth]{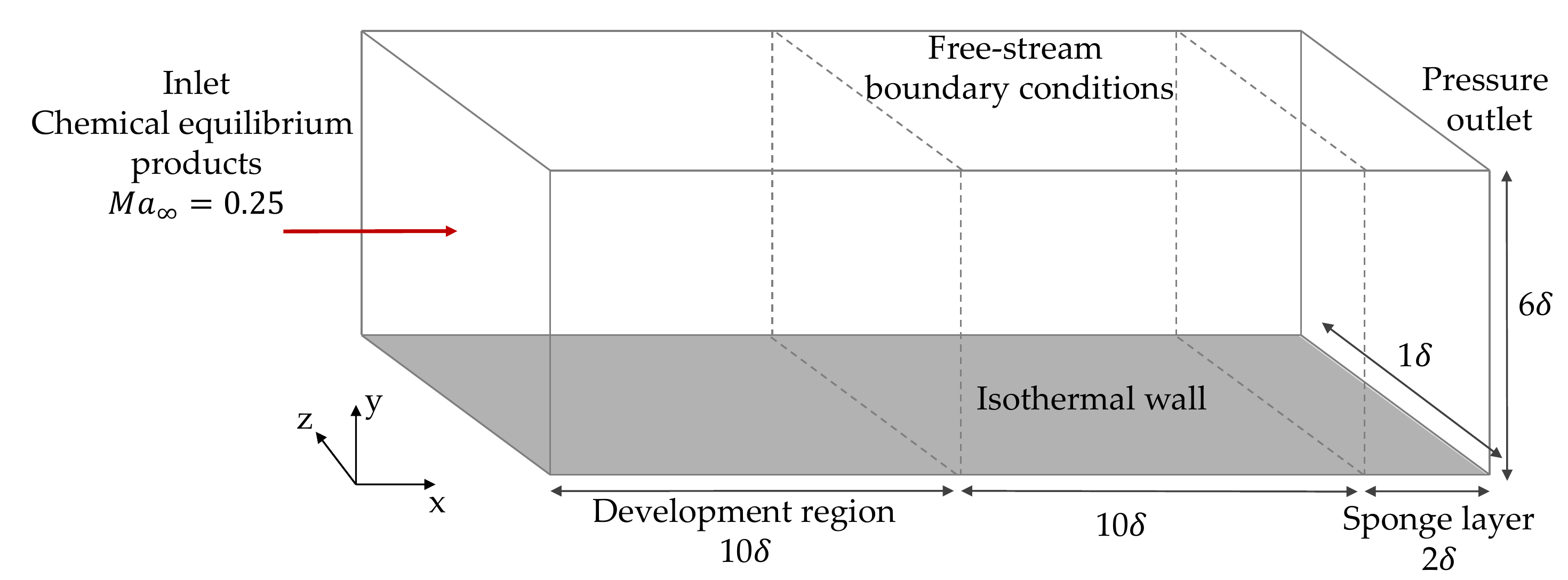}
\caption{\label{fig:DNS_domain} Computational setup for the DNS.  }
\end{figure}

In order to investigate the occurrence of recombination reactions in reacting boundary layers, direct numerical simulations of spatially evolving turbulent boundary layer over an isothermal flat plate are performed. The computational setup consists of a three-dimensional, spanwise periodic domain, which is schematically illustrated in Fig. \ref{fig:DNS_domain}.

For the thermodynamic state at the inlet, the chemical equilibrium composition and temperature for a methane/oxygen mixture with mass mixture ratio equal to 3.0 is chosen, whereas the Mach number is 0.25 and the pressure is 20 bar. This represents the typical composition and velocity conditions found downstream of the main reaction zone in sub-scale rocket engines \citep{perakis2019heat}. A boundary layer thickness of $\delta= 1$ mm is chosen, leading to a Reynolds number $Re_{\delta} = \rho_{\infty} u_{\infty} \delta /\mu_{\infty}= 4678$. The inlet velocity $u_{\infty}$ is 327 m/s, while $\rho_{\infty}$ and $\mu_{\infty}$ are calculated from the chemical equilibrium thermochemical state and are equal to 1.41 kg/m$^3$ and 9.87$\times 10^{-5}\,\mathrm{Pa s}$ respectively.

As described in Xu et al. \cite{xu2004assessment}, the specification of appropriate methods to generate inflow boundary conditions for LES and DNS of compressible boundary layers is important for the development of the turbulence in the domain. In spatially evolving boundary layers, the most convenient procedure involves the definition of a laminar profile with disturbances far upstream, allowing for a transition to turbulence. This approach is typically used to investigate transition \cite{rai1993direct,bhaganagar2002direct,ducros1996large} and is not generally applicable to turbulence simulations as it is very costly especially when coupled to the resolution of chemical reactions. It was hence not considered. Instead, more cost-efficient methods include the recycling of time series of instantaneous velocity planes from an auxiliary simulation similar to the work of Li et al. \cite{li2000inflow}, the specification of the inflow by superposition of random fluctuations on mean flows as done by Lee et al.  \cite{lee1992simulation} and the parallel-flow boundary layer method used by Lund et al. \cite{lund1994large,lund1996large}. In practice, the inflow boundary typically has to be displaced upstream of the region of interest in order to allow for relaxation of the errors made in approximating the inflow conditions. The inclusion of such a "development section" adds to the overall cost of the simulation and therefore one would like to minimize its extent while at the same time, trying to minimize the cost associated with generating the inflow data themselves.

In the present work, the velocity inflow profile is prescribed by a synthetic turbulent flow, of which the mean flow obeys the law of the wall \citep{pope2001turbulent} with boundary-layer thickness of $\delta=1\,$mm. The turbulent perturbations with turbulence intensity of 0.2 in the freestream are generated using the method of Klein et al. \cite{klein2003digital}. This method is able to generate pseudoturbulent inflow conditions based on digital filtering of random data and is able to reproduce prescribed second order statistics as well as auto-correlation functions. It hence provides advantages over the classical approach of using random fluctuations, by reducing the long development section which is induced due to the lack of proper phase information and non-linear energy transfer in random methods. 

Based on these considerations and on typical development lengths reported for inflow turbulence \cite{lund1998generation,le1993direct}, a development region of 10 boundary layer thicknesses is considered. Also, a sponge layer is applied before the outlet to suppress any numerical wave propagation. In the following analysis, $x=0$ represents the start of the domain after the development region and not the inflow boundary.

The mesh in wall-normal distance is discretized using 236 grid points following a geometric growth rate with a first-cell height corresponding to $y^+\approx 0.3$ and adequate points to resolve the viscous sublayer following the recommendations in Moser et al. \citep{moser1999direct}. The mesh in the streamwise and spanwise directions consists of 1400 and 80 equidistant points, respectively. This corresponds to grid resolutions of $\Delta x^+ \approx 11$ and $\Delta z^+ \approx 12$ in the area of interest (after the development region).

\begin{figure}
\includegraphics[width=.43\textwidth]{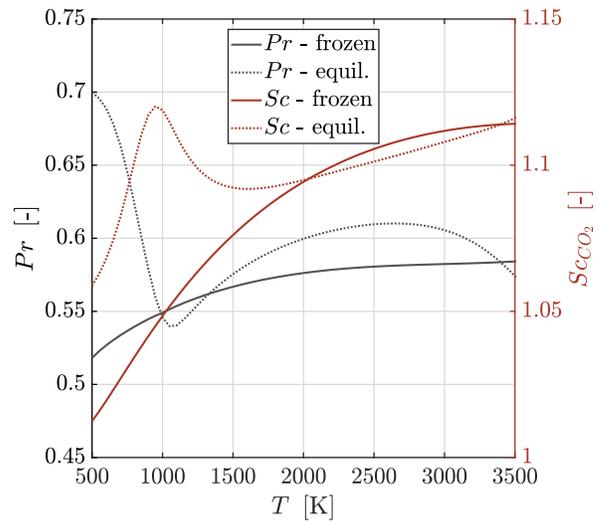}
\caption{\label{fig:Pr_Sc} Prandtl and Schmidt numbers as a function of temperature for the mixture at 20 bar and O/F = 3.0. }
\end{figure}

For locations further away from the wall, the flow is dominated by energy-transferring motions and there, the grid resolution is better evaluated in terms of the Kolmogorov length scale $\eta_K = ((\mu/\rho)^3 \rho/\epsilon)^{1/4}$, where $\epsilon$ is the dissipation rate. Due to the presence of heat and mass transfer, apart from the Kolmogorov scale, the thermal Batchelor scale $\eta_T = \eta_K / \sqrt{Pr}$ \cite{batchelor1959small} and scalar Batchelor scale $\eta_S = \eta_K / \sqrt{Sc}$ \cite{schwertfirm2007dns} have to be resolved as well. As the Prandtl $Pr$ and Schmidt $Sc$ numbers of the examined mixture are close to unity for the entire temperature range, $\eta_K$, $\eta_T$ and $\eta_S$ are all on the same order of magnitude. The variations of the non-dimensional numbers for the mixture are shown in Fig. \ref{fig:Pr_Sc} for the two extreme cases of frozen chemistry and equilibrium conditions. The Schmidt number of CO$_2$ is chosen as it has the lowest diffusivity among the dominant reaction species. The values are within the resolution requirements of $\Delta x < 12 \eta_T$, $\Delta y < 2 \eta_T$ and $\Delta z < 6 \eta_T$, as also reported in other DNS studies \citep{zonta2012modulation,lee2013effect,patel2017scalar}.

\begin{figure}
\includegraphics[width=.42\textwidth]{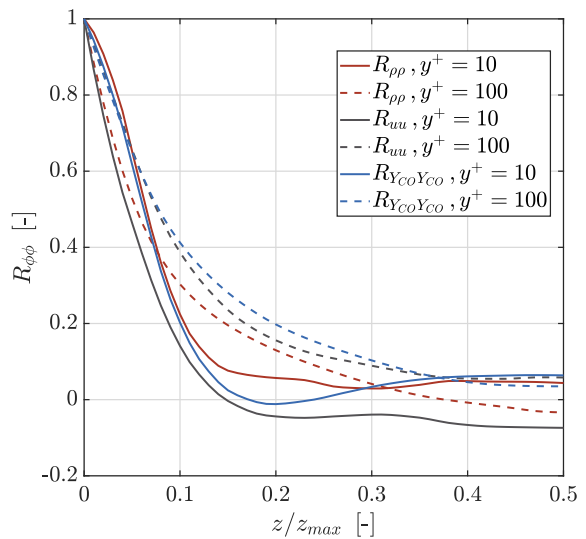}
\caption{\label{fig:2pntCorr} Two-point correlation for downstream position $x/\delta =5$.  }
\end{figure}

The spanwise extent of the domain was chosen to ensure that it exceeds the minimal size of $l_z^+=100$, reported in Jimenez et al. \cite{jimenez1991minimal}. For a periodic boundary condition to be reasonably accurate the two-point correlations are required to be close to zero at a distance of half the domain size. In Fig. \ref{fig:2pntCorr}, we show the two-point correlations for density ($R_{\rho \rho}$), streamwise velocity ($R_{u u}$) and CO mass fractions ($R_{Y_{CO} Y_{CO}}$) at wall-normal positions $y^+=10$ and $y^+=1000$ and streamwise location $x/\delta=5$. 

For the simulations presented in the following sections, the governing equations for continuity, momentum, energy and transported scalars are solved as follows:

\begin{equation}
\partial_{t} \rho+\nabla \cdot(\rho \boldsymbol{u})=0
\label{eq:continuity}
\end{equation}

\begin{equation}
\partial_{t}(\rho \boldsymbol{u})+\nabla \cdot(\rho \boldsymbol{u} \boldsymbol{u}+p \boldsymbol{I})=\nabla \cdot \boldsymbol{\tau}
\label{eq:momentum}
\end{equation}

\begin{equation}
\partial_{t}(\rho E)+\nabla \cdot[\boldsymbol{u}(\rho E+p)]=\nabla \cdot(\boldsymbol{\tau} \cdot \boldsymbol{u})-\nabla \cdot \dot{\bf{q}}
\label{eq:energy}
\end{equation}

\begin{equation}
\partial_{t}(\rho Y_k)+\nabla \cdot(\boldsymbol{u}\rho Y_k)=-\nabla \cdot \boldsymbol{j}_k + \dot{\omega}_k M_k
\label{eq:species}
\end{equation}

\noindent where $\rho$ is the density, $\boldsymbol{u}$ is the velocity vector, $p$ is the pressure, $\boldsymbol{\tau}$ is the viscous stress tensor, $\boldsymbol{\dot{q}}$ is the heat flux vector and $E$ is the specific total energy defined as the sum of the specific internal energy $e$ and the kinetic energy $ \left| \boldsymbol{u} \right|^2/2$. The ideal gas equation of state is used as closure for the system of equations. 

The viscous stress tensor, heat flux and species diffusion terms are given by:

\begin{equation}
\boldsymbol{\tau}=\mu\left[\nabla \boldsymbol{u}+(\nabla \boldsymbol{u})^{T}\right]-\frac{2}{3} \mu(\nabla \cdot \boldsymbol{u}) \boldsymbol{I} \; ,
\end{equation}

\begin{equation}
\boldsymbol{q}=-\lambda \nabla T \; ,
\end{equation}

\begin{equation}
\boldsymbol{j}_k = -\rho\left( D_{k} \frac{M_{k}}{M} \nabla X_{k} -Y_{k} \sum_{l=1}^{N} D_{l} \frac{M_{l}}{M} \nabla X_{l} \right) \; ,
\end{equation}

To account for the chemical reactions and the production term $\dot{\omega}_k$  the present DNS study considers the detailed GRI 3.0 chemical mechanism \cite{SmithGRIMech}. In the case of the inert simulations, the chemical source term is set to zero.

The system of equations is discretized based on a finite-volume approach and a high-order non-dissipative scheme is used for the convective flux discretization \citep{ma2017entropy}, which is fourth-order accurate on uniform meshes. A strong stability-preserving third-order Runge-Kutta scheme  is used for time advancement. \cite{gottlieb2001strong}.
A Strang-splitting scheme \cite{strang1968construction,wu2019efficient} is employed to separate the convection operator from the remaining operators of the system. A sensor-based hybrid central-ENO scheme is used to capture flows with large density gradients and to minimize the numerical dissipation while stabilizing the simulation. For regions where the density ratio between the reconstructed face value and the neighboring cells exceeds 25\%, a second-order ENO reconstruction is used on the left- and right-biased face values, followed by an HLLC Riemann flux evaluation. The flow is well resolved and the ENO scheme is active on less than 0.01\% of all cell faces.

\section{Results}

The results of the DNS are presented in this section for four different computational setups. Reacting and inert simulations are carried out for two different values of the wall temperature (500$\,$K and 1000$\,$K) resulting in a total of four cases. In this section will mainly focus on the results of the case with 500$\,$K wall temperature and we only consider the 1000$\,$K case to show qualitative differences.

\subsection{Velocity scaling}

For constant-property wall-bounded turbulent flows, the near-wall time-averaged velocity follows the law of the wall \cite{pope2001turbulent} with the wall-normalized velocity given by $u^{+}=y^{+}$ in the viscous sublayer ($y^+<5$) and by the logarithimic relationship $u^{+}=\frac{1}{\kappa} \log \left(y^{+}\right)+B$ in the logarithimic layer ($30 \leq y^+\leq 0.1 \delta$).

For variable-property flows, both the fluid density and the dynamic viscosity are functions of temperature and pressure, leading to additional complexities. One commonly used velocity transformation that allows the scaled velocity to collapse with the incompressible law of the wall is the van Driest transformation \cite{van1951turbulent}: 

\begin{equation}u_{\mathrm{VD}}^{+}=\int_{0}^{u^{+}}\left(\frac{\bar{\rho}}{\bar{\rho}_{w}}\right)^{1 / 2} d u^{+} \; .
\label{eq:VanDriest}
\end{equation}

\begin{figure}
\includegraphics[width=.42\textwidth]{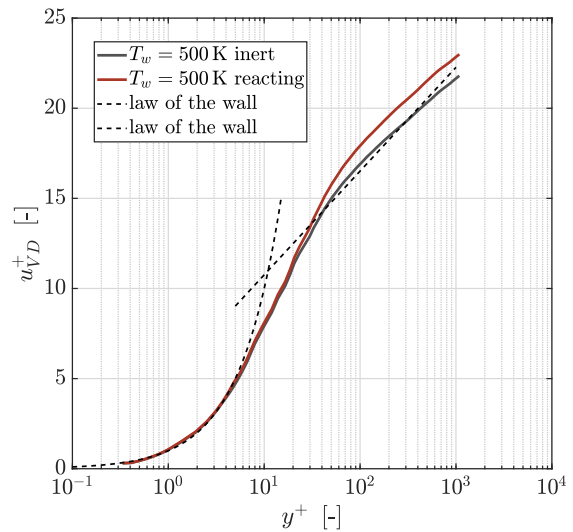}
\caption{\label{fig:VanDriest} Normalized velocity profile of the inert and reacting DNS using the van Driest transformation.}
\end{figure}

A comparison of the velocity profiles from the DNS is shown in Fig. \ref{fig:VanDriest}, where the data  support the validity of the van Driest transformation 
for compressible reacting flows. This can be attributed to the  density fluctuations in compressible low-Mach number flows, arising from mean fluid-property variations. This confirms the 
Morkovin hypothesis \citep{morkovin1962effects}, which has been used in the past to investigate the compressible turbulent boundary layer in the same line as for the incompressible one by accounting for the effects of mean density variations \cite{so1998morkovin}. In the range of Mach numbers explored in the present study, our DNS results support previous experimental and numerical studies that showed the validity of the van Driest transformation in scaling the velocity profile with wall heat transfer and even chemical reactions \citep{huang1993skin,so1994logarithmic,morinishi2004direct,wang1996large,cabrit2009direct}.
In the presence of higher Mach-number flows and real-fluid effects \citep{ma2018structure}, it is expected that additional effects require consideration, related to the compressible nature of the fluid.

\subsection{Temperature and species mass fractions}

\begin{figure}
\includegraphics[width=.4\textwidth]{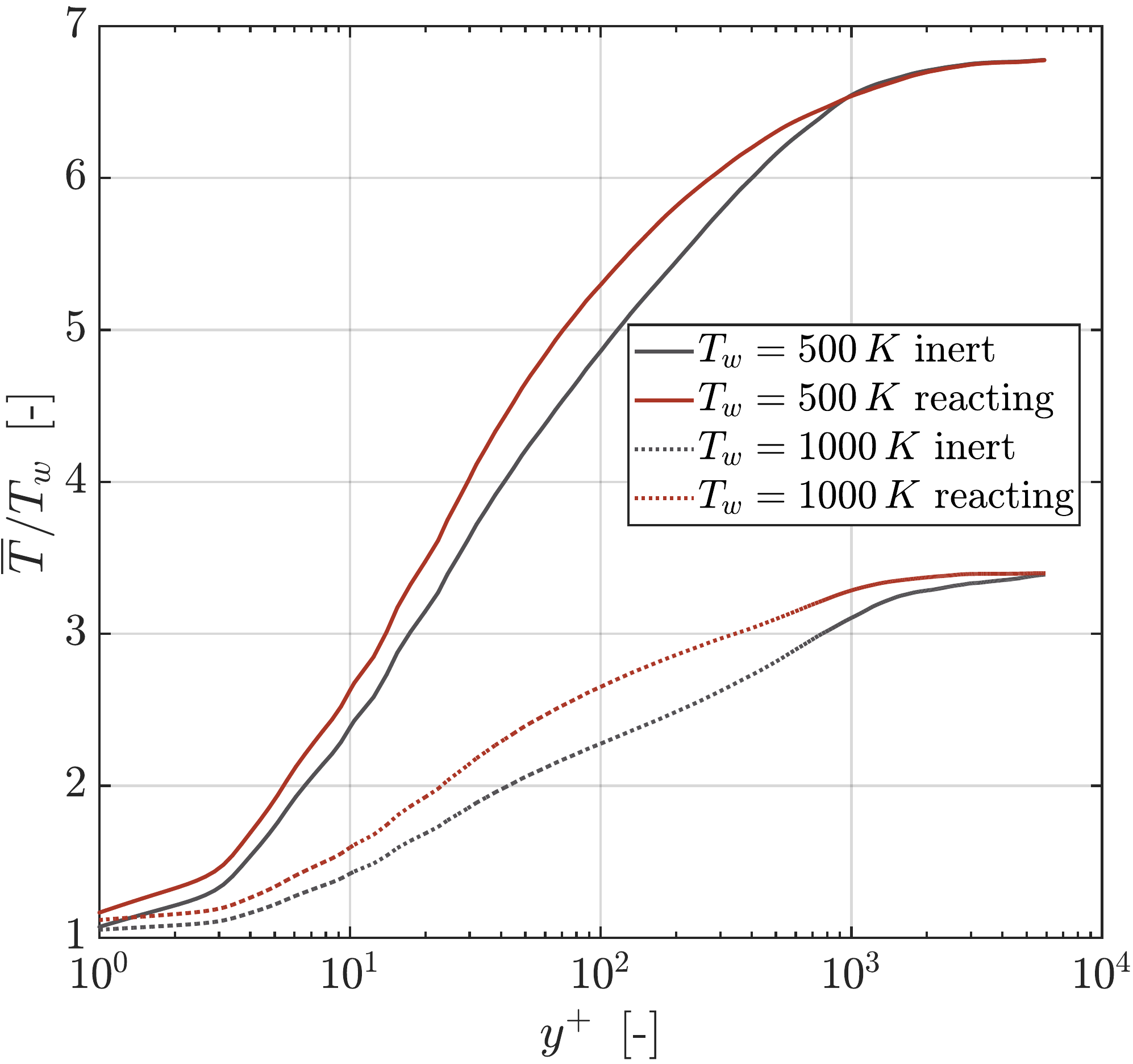}
\caption{\label{fig:Temperature}Temperature profile within the turbulent boundary layer.}
\end{figure}

The mean temperature profiles (normalized by the wall temperature) are shown as a function of wall-normal distance in Fig. \ref{fig:Temperature}. The effect of chemical reactions in the boundary layer is evident for both wall temperatures. Specifically, the temperature profiles for the reacting cases show higher values in the range $10 \leq y^+ \leq 1000$. The increased temperature is a result of the exothermic reactions that are induced in the low-enthalpy boundary-layer region.

\begin{figure*}
(a)
\includegraphics[width=.4\textwidth]{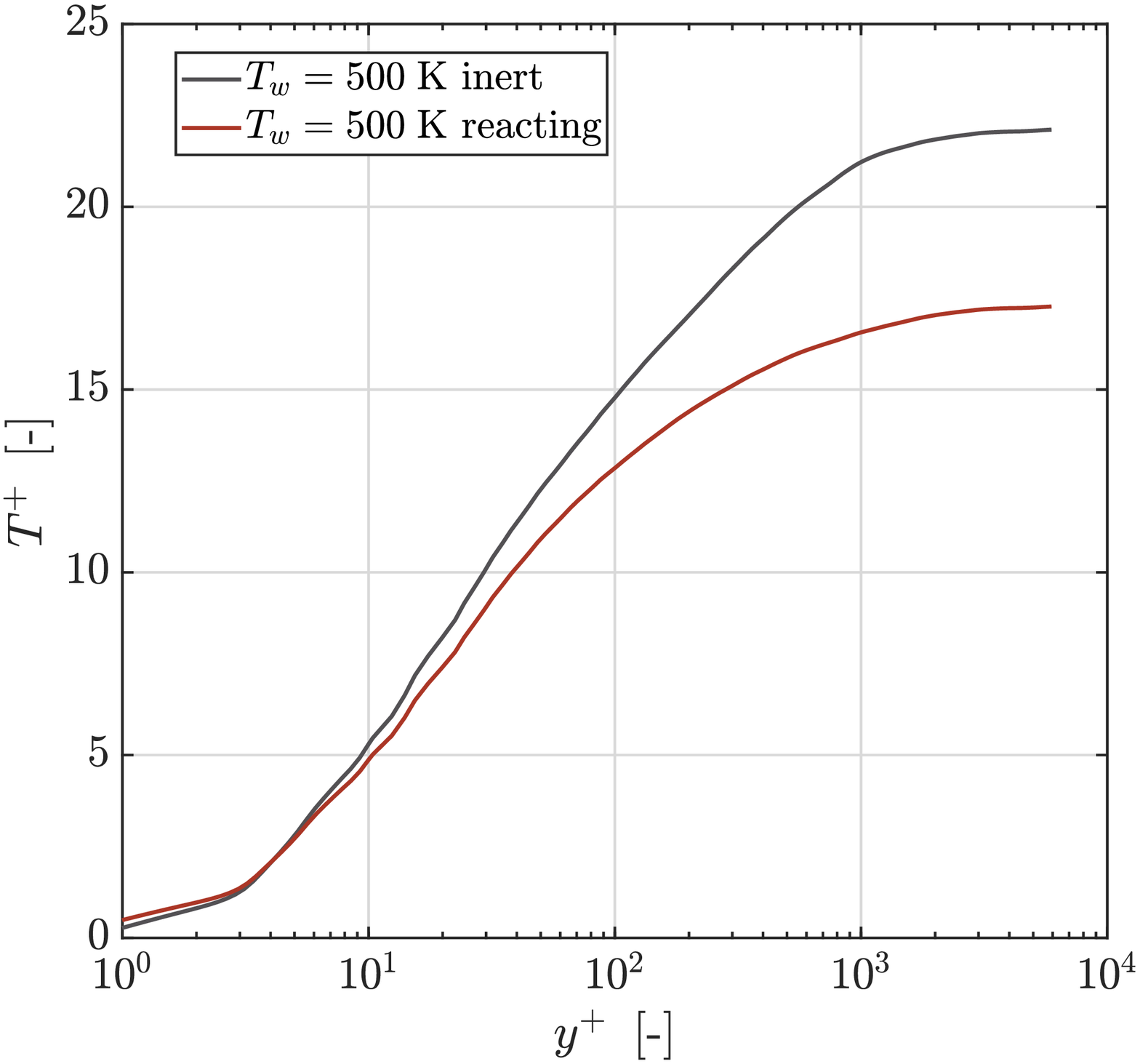}
(b)
\includegraphics[width=.4\textwidth]{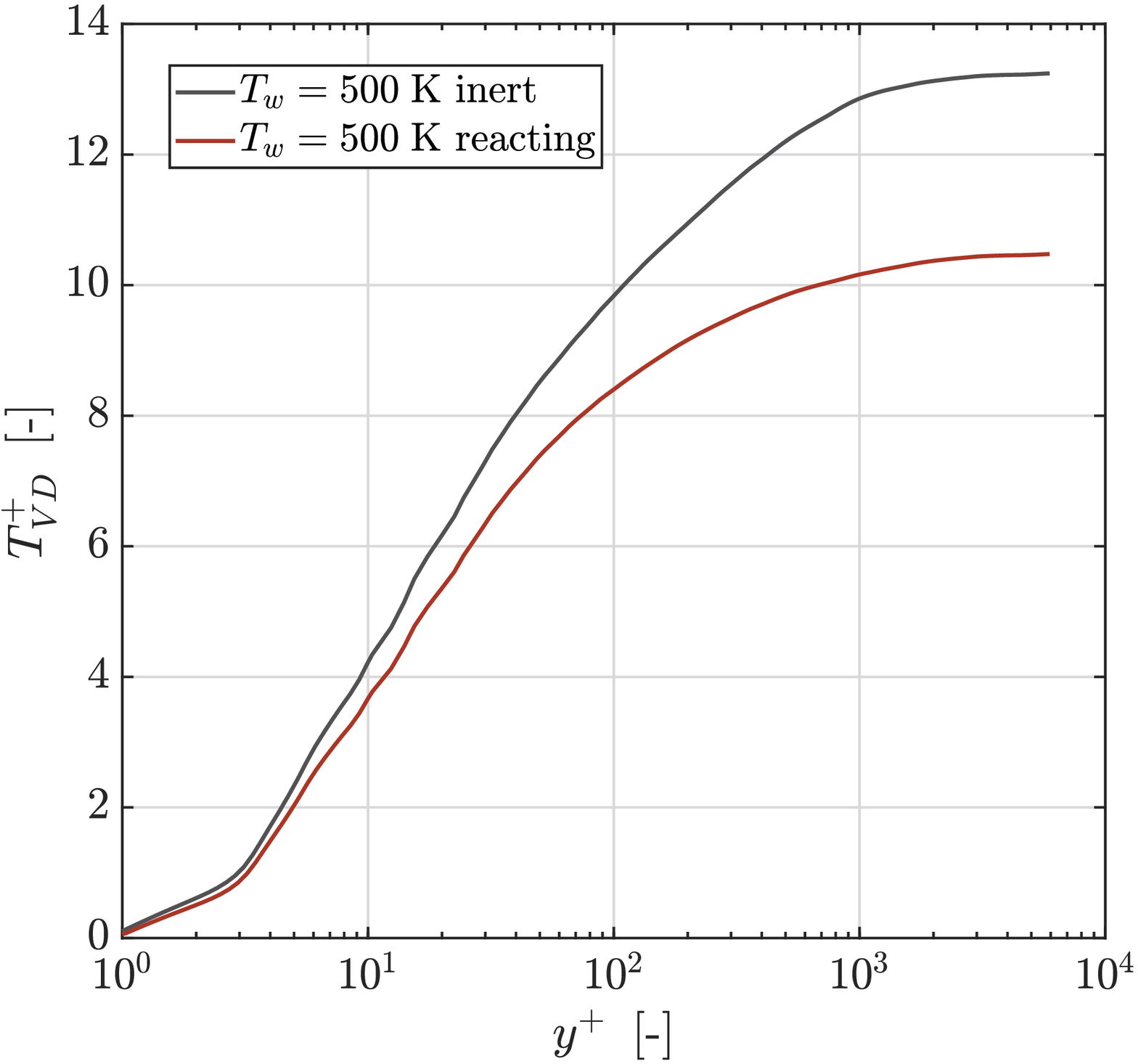}
\caption{\label{fig:TemperaturePlus}Profiles of the (a) transformed temperature $T^+$ and (b) van Driest scaled temperature $T^+_{VD}$ within the turbulent boundary layer.}
\end{figure*}

Fig. \ref{fig:TemperaturePlus}(a)  shows the profile of the mean transformed temperature $T^+$, which is defined as the temperature difference $\theta=\overline{T}-T_{w}$ normalized by the friction temperature $T_{\tau} = \overline{q_w}/\left( \overline{\rho_w} \, \overline{c_{p}} _{,w} u_{\tau} \right)$.

\begin{equation} T^+ = \frac{\overline{T}-T_{w}}{T_{\tau}} \; .
\label{eq:TempScaling}
\end{equation}

The discrepancies observed between the reacting and inert cases indicate that the wall heat flux is sensitive to chemistry. Specifically, the values for the inert case are higher than the reacting case, indicating the presence of an appreciable heat transfer augmentation due to the chemical reactions.

Using the van Driest transformed temperature profile \citep{patel2017scalar}

\begin{equation}T_{\mathrm{VD}}^{+}=\int_{0}^{T^{+}}\left(\frac{\bar{\rho}}{\bar{\rho}_{w}}\right)^{1 / 2} d T^{+} \; ,
\label{eq:VanDriestT}
\end{equation}

\noindent as shown in Fig. \ref{fig:TemperaturePlus}(b), does not lead to a collapse of the inert and reacting profiles. Compared to the findings in Fig. \ref{fig:VanDriest}, where the velocity profiles for the reacting and inert cases showed similar properties, it is clear that the chemical reactions have an important effect on the fluxes at the wall, even if they do not influence the mean velocity profiles. This finding aligns with the results reported by Carbit et al. \citep{cabrit2009direct} and justifies the need of taking care of fluid heterogeneity in wall models.

\begin{figure}
\includegraphics[width=.48\textwidth]{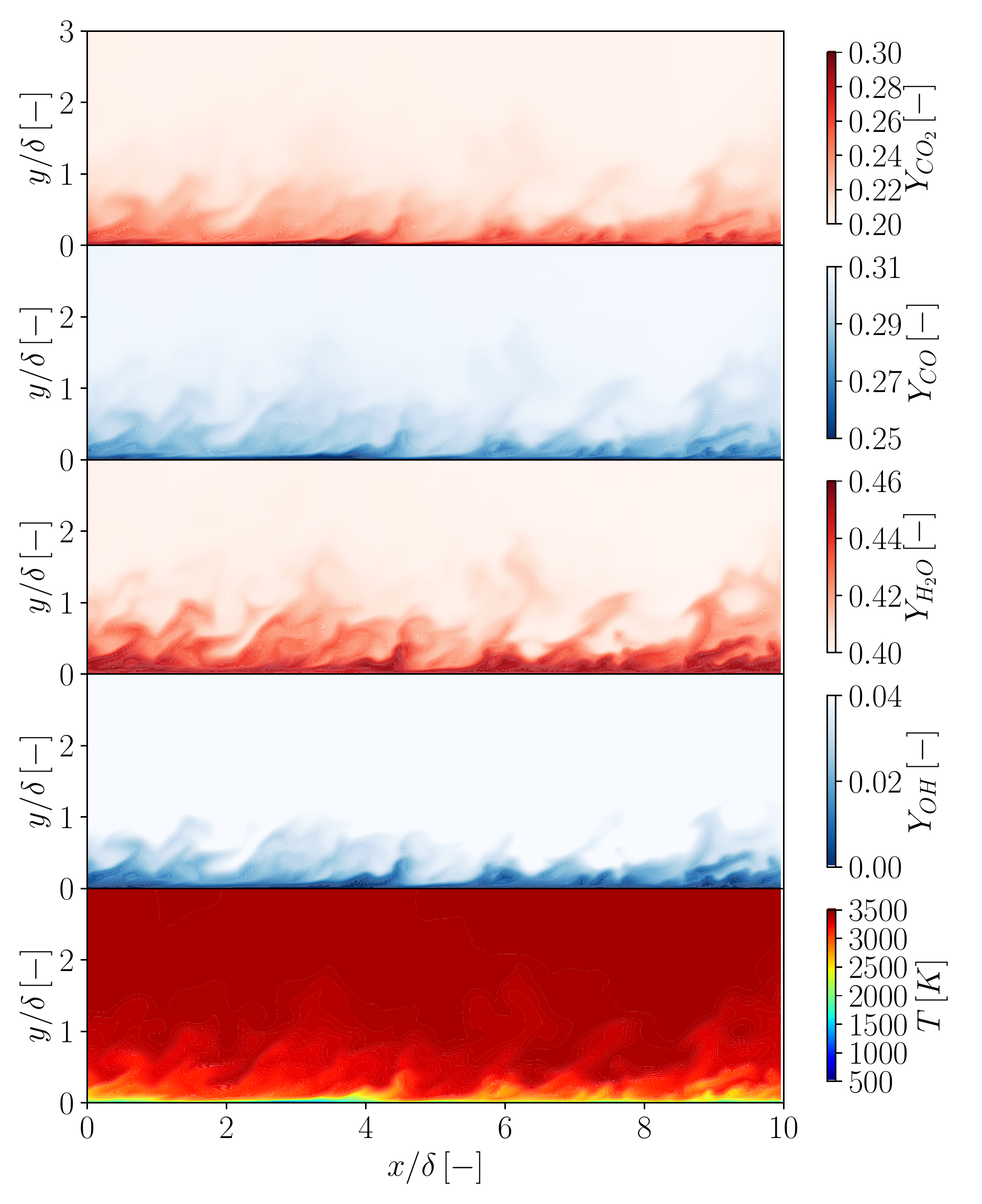}
\caption{\label{fig:instant2D_500K}Instantaneous species mass fractions and temperature field for the reacting case with wall temperature equal to 500 K.}
\end{figure}

To illustrate the effect of the chemical reactions, the instantaneous fields of major species and temperature are shown in Fig. \ref{fig:instant2D_500K}. The formation of a distinctive species boundary layer in the vicinity of the cooled wall is evident. Specifically, the recombination of CO to CO$_2$ and that of OH to H$_2$O is prominent throughout the entire domain. Although only major species are shown here, the recombination reactions influence all radicals that also react and form stable species. The locations with increased CO$_2$ mass fraction and hence reduced CO mass fraction are directly correlated to regions with lower temperature.

\begin{figure*}
(a)
\includegraphics[width=.375\textwidth]{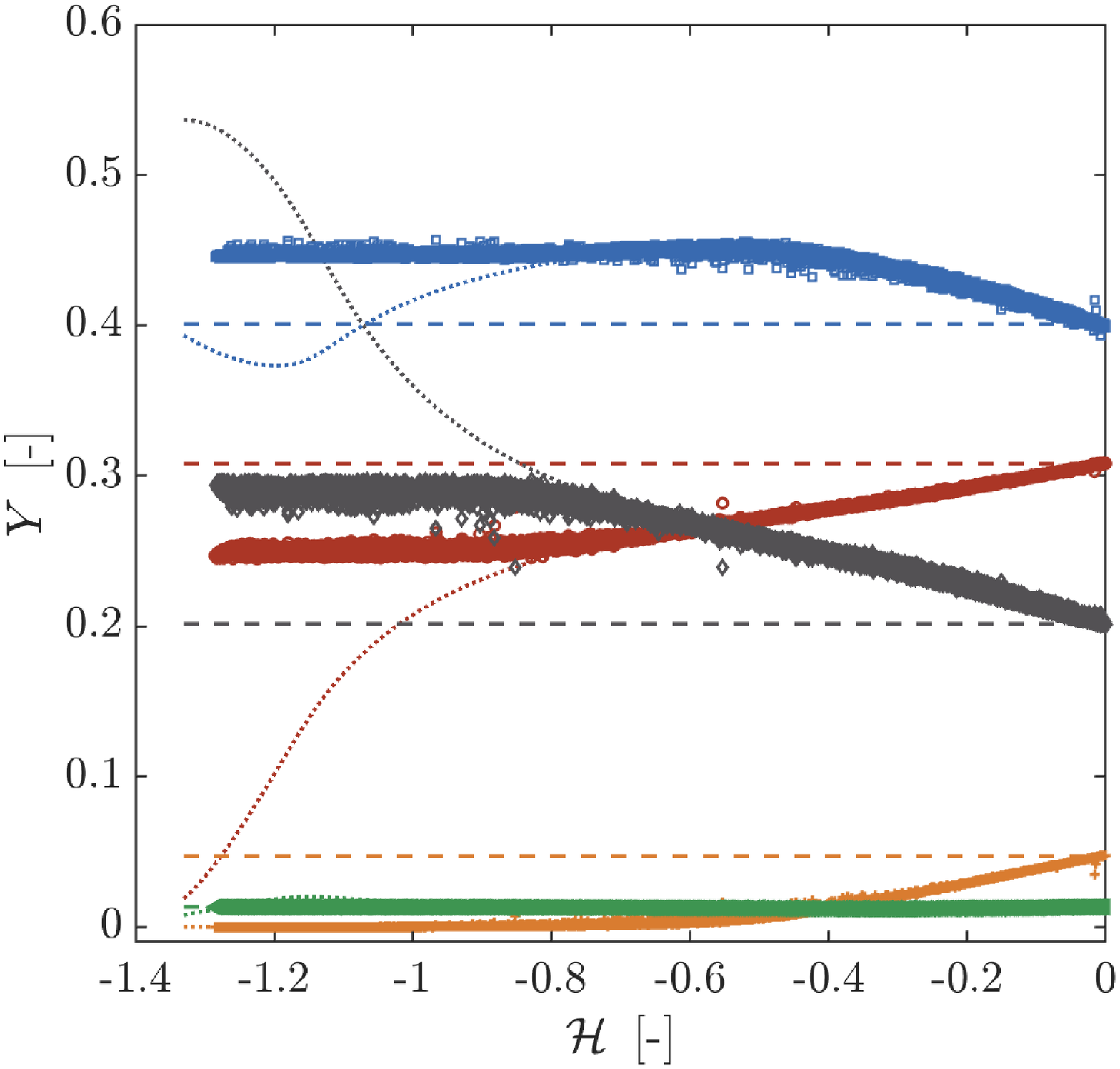}
(b)
\includegraphics[width=.341\textwidth]{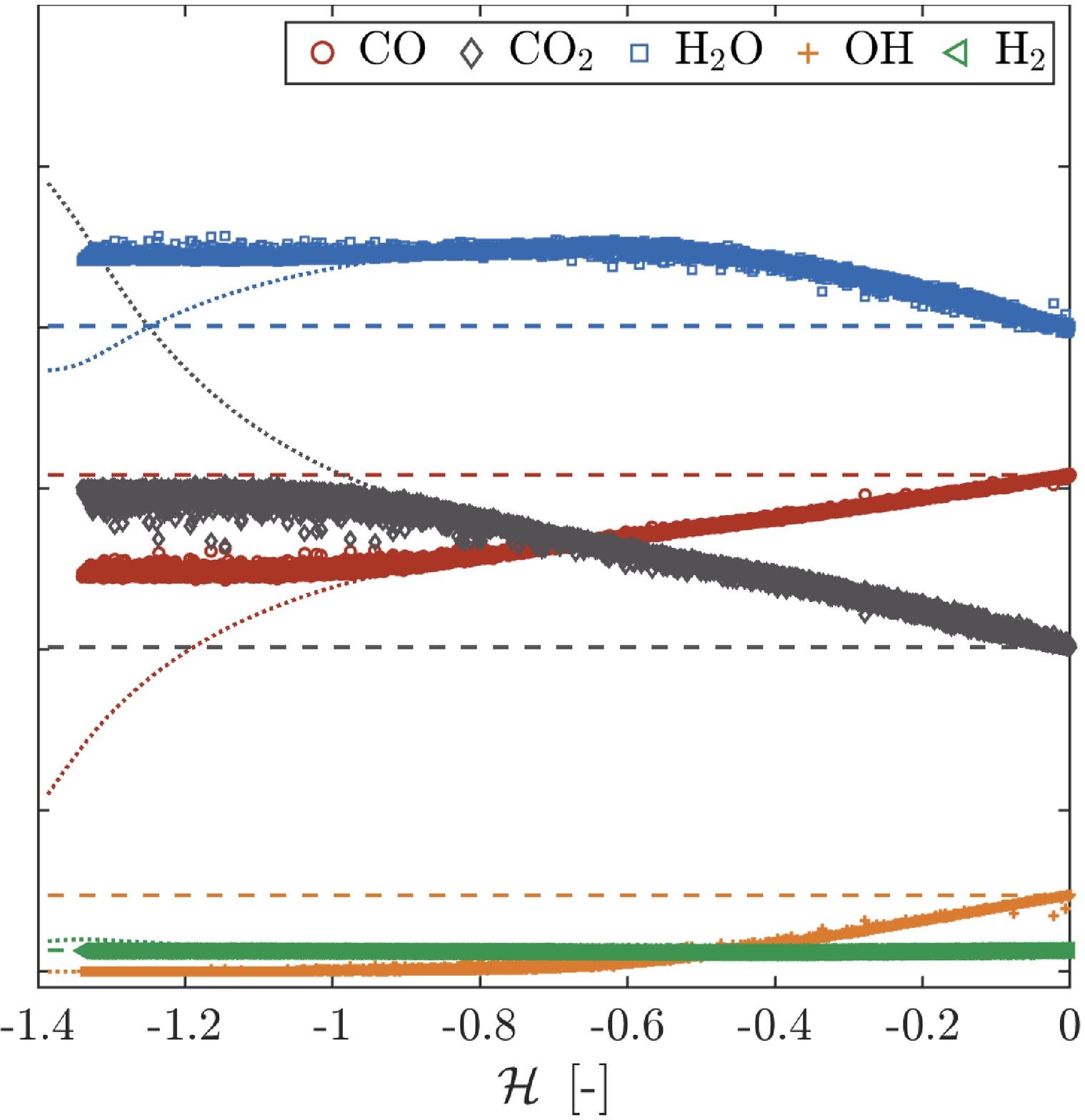}
\caption{\label{fig:Scatter} Instantaneous enthalpy conditioned mass fractions for wall temperature equal to (a) 500 K and (b) 1000 K. Dotted lines correspond to equilibrium mass fractions and dashed lines correspond to frozen mass fractions. }
\end{figure*}

In order to obtain a better understanding of where those reactions occur and to quantify the correlation between species mass fractions and enthalpy, we examine Fig. \ref{fig:Scatter}, which illustrates the distribution of the major species as a function of the normalized enthalpy loss. $\mathcal{H}$, which is defined as:

\begin{equation}
\mathcal{H} = \frac{h-h_{ad}}{h_{wall} - h_{ad}} \; ,
\label{eq:Enthalpy_loss}
\end{equation}

\noindent and lower values of $\mathcal{H}$ correspond to conditions closer to the wall. $h_{ad}$ represents the adiabatic enthalpy of the mixture, i.e. the enthalpy of the hot products at the inlet and $h_{wall}$ the enthalpy that the gas would have when cooled down to the wall temperature without changing its composition.

The scatter plots in Fig. \ref{fig:Scatter} show a clear correlation between the normalized enthalpy loss and the species concentration. Compared to the adiabatic values ($\mathcal{H} =0$), a decrease in enthalpy appears to promote the recombination of OH to H$_2$O and of CO to CO$_2$. A noteworthy remark is the small dispersion of the instantaneous data in the enthalpy space, as a very small variation is observed with the point cloud being concentrated around a well established average for each species. We compare the change in concentration with the theoretical chemical equilibrium for each enthalpy level, which is illustrated by the dotted lines in Fig. \ref{fig:Scatter}. For enthalpy losses $\mathcal{H} \geq -0.8 $ the results follow the chemical equilibrium concentration, indicating that the reaction rates are faster than the competing physical phenomena. In lower enthalpy environment, however, the deviation increases, which implies that the rate of change of the gas composition is now dictated by larger time-scales. Finally, the shift towards the recombined products is not continuously increasing with lower enthalpy but rather reaches a plateau, indicating a freezing of the reactions directly at the wall. 

It is noteworthy that the minimum value of $\mathcal{H}$ in the domain is smaller than -1, which corresponds to a quenched (non-reacting) mixture that is cooled to the wall temperature. This arises from the fact that the major combustion products H$_2$O and CO$_2$ have a lower formation enthalpy and a lower specific heat capacity than OH and CO. As the gas approaches the wall, the heat loss favors the formation of stable products and results in a lower enthalpy compared to its frozen equivalent.

\begin{figure}
\includegraphics[width=.4\textwidth]{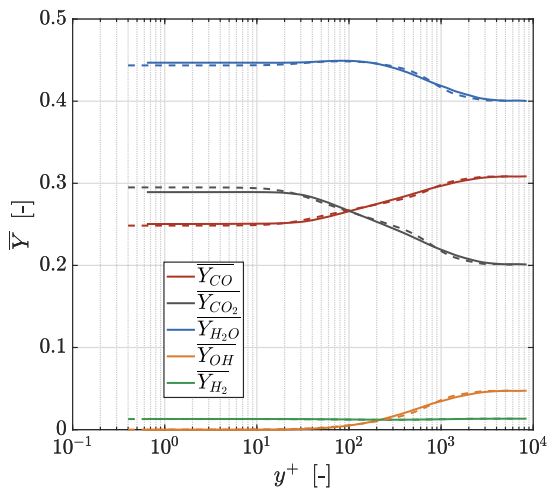}
\caption{\label{fig:SpeciesYplus} Mass fractions of the major species as a function of the wall-normal distance for the reacting cases with $T_w=500\,$K (solid line) and $T_w=1000\,$K (dashed line).}
\end{figure}

The spatial distribution of the species composition is illustrated in Fig. \ref{fig:SpeciesYplus}, where the average mass fraction of the major species is plotted as a function of the wall-normal distance for the reacting cases with $T_w=500\,$K and $T_w=1000\,$K.
Distinct regions can be identified when examining the species mass fraction profiles. Starting from locations away from the wall ($y^+\geq 3000$), it appears that no significant change occurs in the composition of the gas, leading to non-varying mass fractions. However at $y^+\approx 3000$, a species boundary layer starts to develop. The conversion of CO to CO$_2$ is observed, along with the recombination of OH to H$_2$O. The mass fraction of H$_2$ on the other hand is unaltered. For both wall temperatures, the normalized wall-distance at which the chemical reactions take effect are comparable. In accordance with the enthalpy conditioned diagram in Fig. \ref{fig:Scatter} which provides information about a "freezing" of the reactions in enthalpy space, a similar termination of the chemical recombinations is evident in physical space as well. For wall-normal distances with $y^+<10$, all major species have reached a constant value, which infers a suppression of the reaction rates because of the low temperature.


\begin{figure*}
(a)
\includegraphics[width=.375\textwidth]{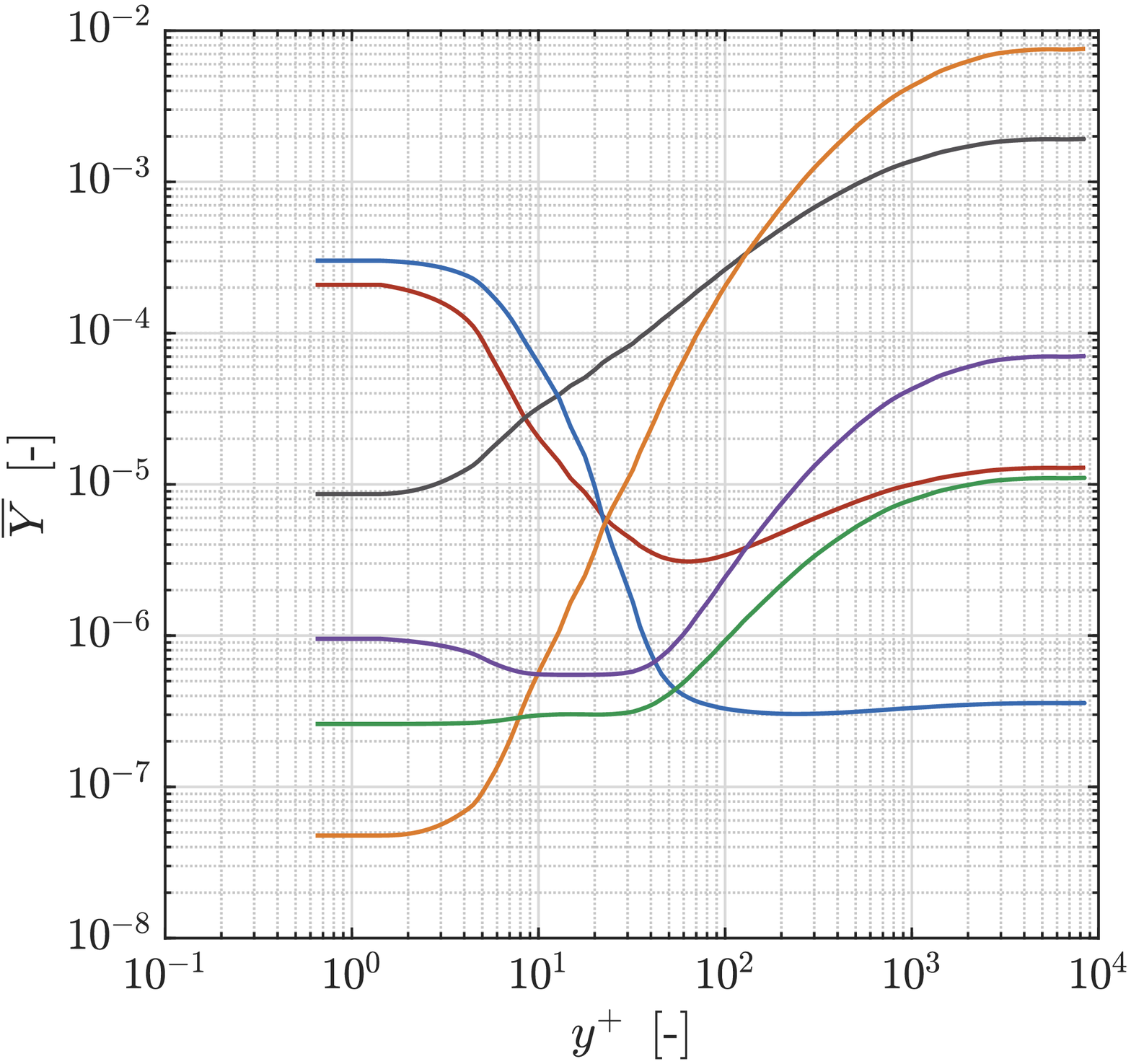}
(b)
\includegraphics[width=.341\textwidth]{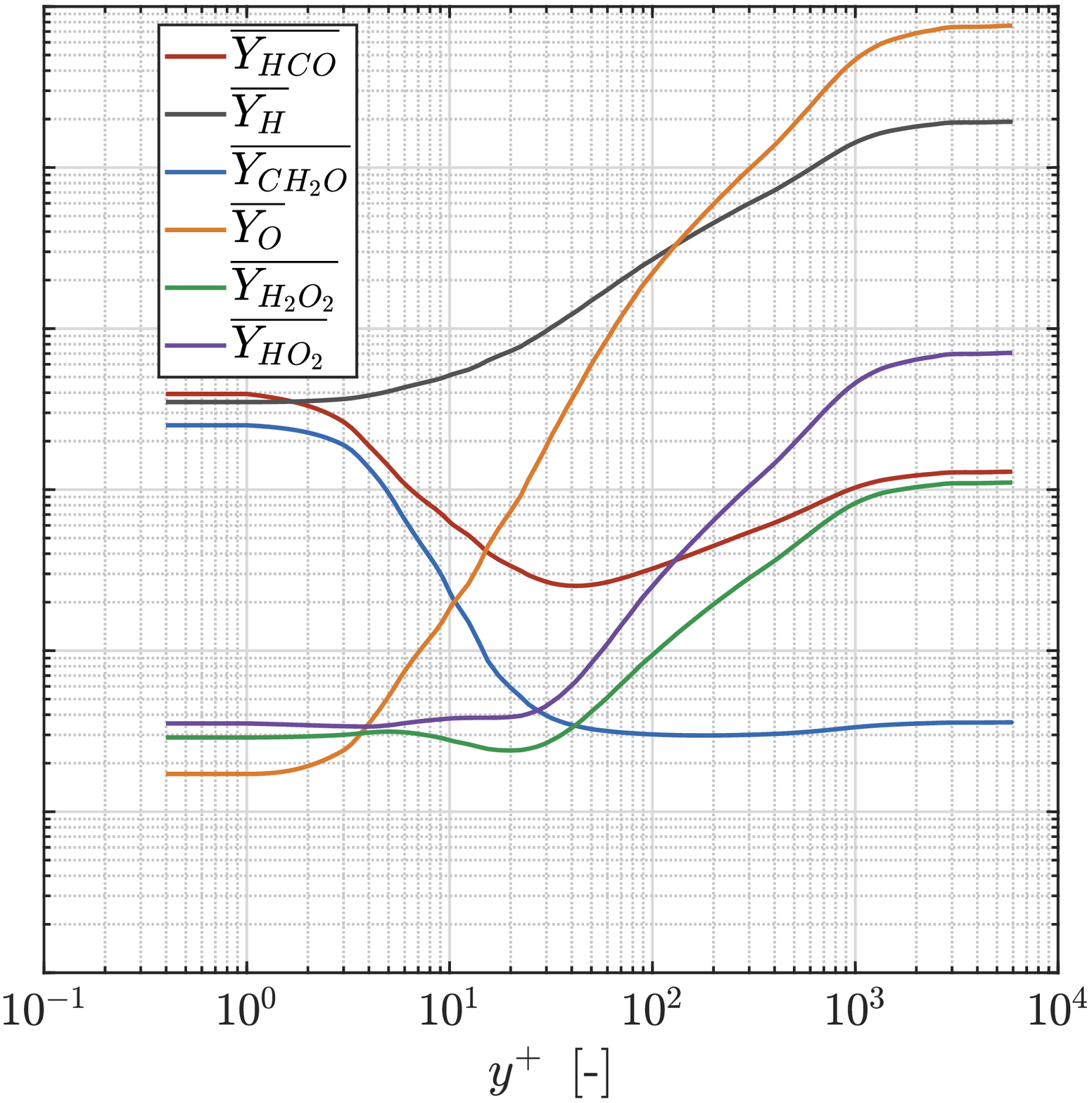}
\caption{\label{fig:SpeciesYplusMinor} Minor and radical species mass fractions for the reacting cases as a function of the wall-normal distance for (a) $T_w=500\,$K and (b) $T_w=1000\,$K.}
\end{figure*}

Apart from the major species, shown in Fig. \ref{fig:SpeciesYplus}, the mass fraction profiles of the radical species are depicted in Fig. \ref{fig:SpeciesYplusMinor}. Similar to the behavior of major species, the radical species concentrations are nearly constant further away from the wall, where the gas is in a state of chemical equilibrium ($y^+>2000$). At locations closer to the isothermal wall, the exothermic recombination reactions result in a reduction in the mass fractions of all radicals. This remains for most of the species until a quenching of the reactions is reached at $y^+<10$. For the formyl radical (HCO) and formaldehyde (CH$_2$O), however, a distinct increase in mass fraction is observed starting at $y^+\approx 100$. A second reaction zone is hence present, in which the formation of HCO and   CH$_2$O is favored.

The results of the species profiles both in conditional and physical space are qualitatively similar for the tow wall temperatures. For that reason, only the case with $T_w= 500\,$K is examined in the remaining sections.

\subsection{Reaction path analysis}

\begin{figure*}
(a)
\includegraphics[width=.375\textwidth]{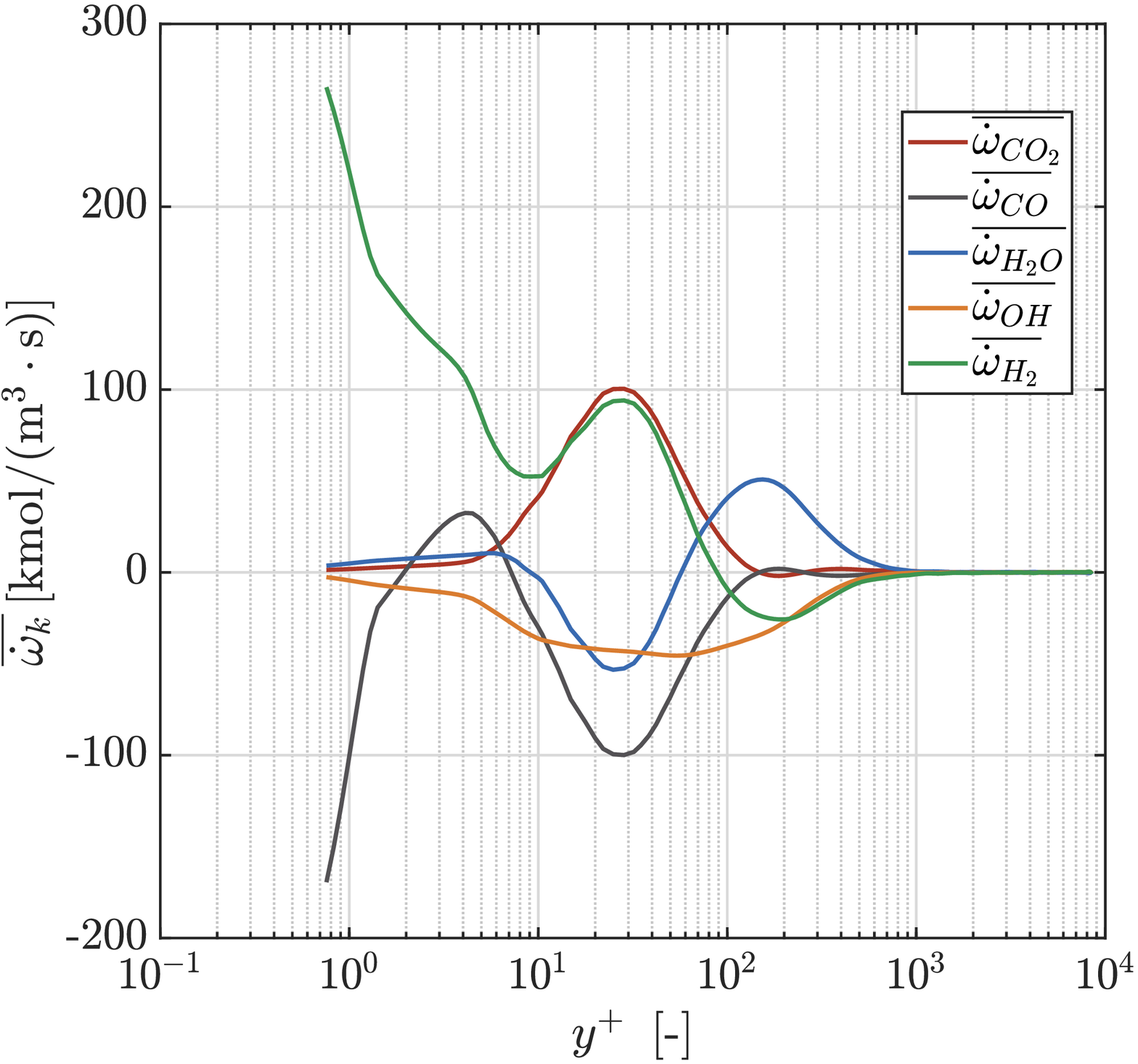}
(b)
\includegraphics[width=.375\textwidth]{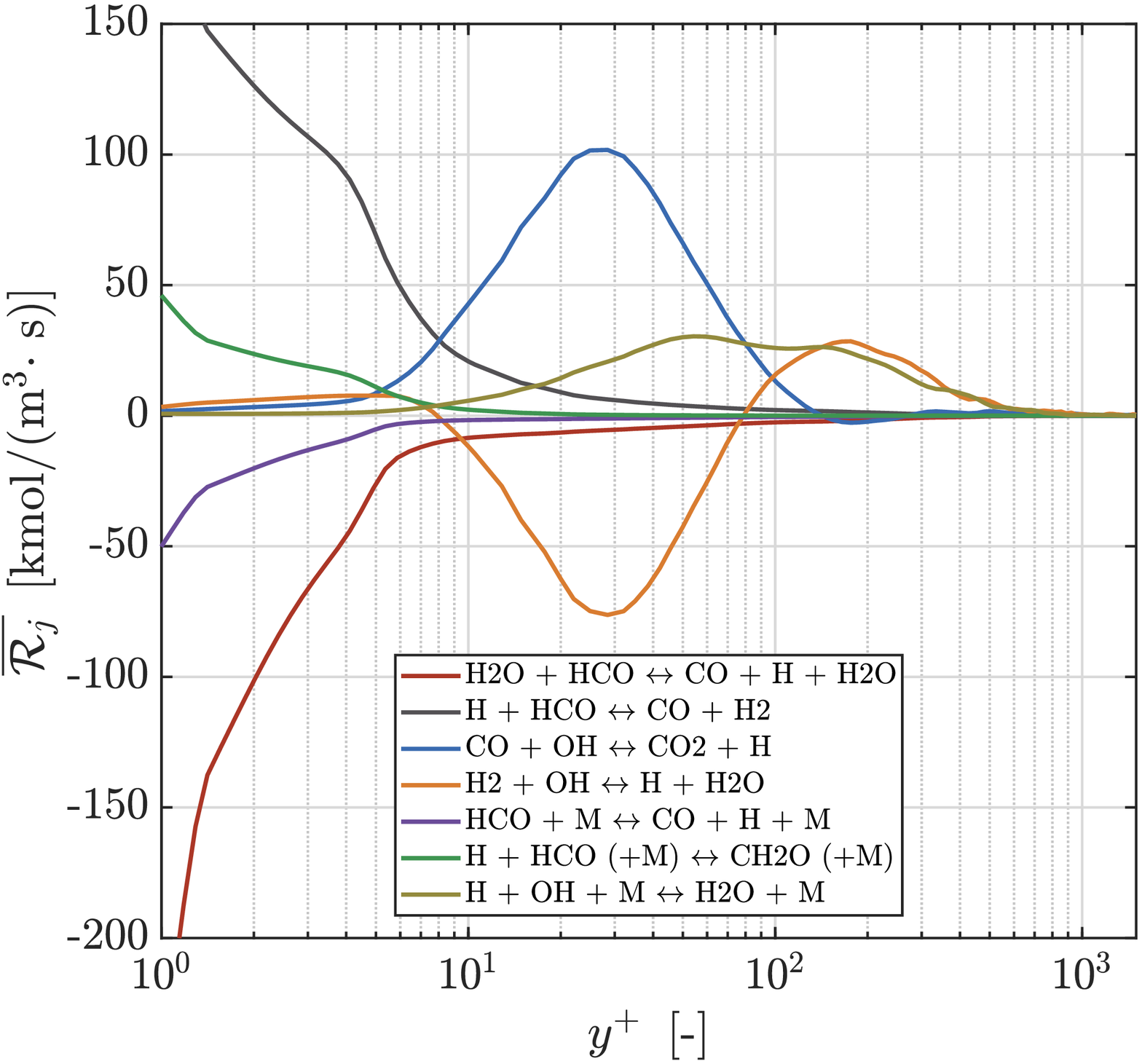}
\caption{\label{fig:OmegaDot} Profiles of (a) mean net chemical production rates of the major species and (b) mean net rates of progress for the most dominant reactions for the reacting case with $T_w=500\,$K.}
\end{figure*}

In order to understand the origin of the recombination reactions, the mean net species molar reaction rates $\dot{\omega}_k$ are plotted in Fig. \ref{fig:OmegaDot}(a). In agreement with the species plots in Fig. \ref{fig:SpeciesYplus}, the reaction rates for $y^+>1000$ are exactly zero, whereas the main reaction zone extends from $y^+\approx 10$ until $y^+\approx 1000$. Starting at locations further from the wall, it is evident that the hydrogen chemistry is activated first, with the reaction rates of H$_2$O, H$_2$ and OH increasing in magnitude already at $y^+\approx 1000$. The carbon chemistry on the other hand (represented here by the reaction rates of CO and CO$_2$) is more localized, with non-negligible values for the reaction rates beginning at $y^+\approx 100$. As far as the net production and destruction of the species is concerned, in the area where the bulk of the reactions takes place (between $y^+\approx 1000$ and $y^+\approx 10$), there are regions with positive net production of water and regions with net consumption of water, with the transition occurring at $y^+\approx 500$. For OH, on the other hand, only net consumption rates are shown. Moreover, as expected based on the scatter results (Fig. \ref{fig:Scatter}), CO$_2$ is predominantly being created in regions where CO has a net consumption rate.

Although the mass fraction of CO appears to be unaffected in regions closer to the wall due to the freezing of the reactions that was established in Fig. \ref{fig:SpeciesYplus}, a sharp increase in the consumption rate of CO is observed for distances $y^+< 10$, with a simultaneous increase in the H$_2$ net creation rate. This effect, which is most prominent for the case with lower wall temperature, occurs at wall-distances that coincide with the regions of increased HCO and CH$_2$O abundance (from Fig. \ref{fig:SpeciesYplusMinor}). This secondary reaction zone is hence activated at low temperatures and is confined within the viscous sub-layer.


In order to investigate the nature of this secondary recombination layer, the reactions with the highest mean net rates of progress $\overline{\mathcal{R}_{j}}$ are explored in Fig. \ref{fig:OmegaDot}(b). The instantaneous rate of progress of reaction $j$ is defined as:

\begin{equation}
    \mathcal{R}_{j} = \Gamma_j \cdot \left( \mathcal{K}_{fj} \prod_{k=1}^{N_{sp}} \mathcal{C}_k ^{\nu_{k j}^{\prime}} - \mathcal{K}_{rj} \prod_{k=1}^{N_{sp}} \mathcal{C}_k ^{\nu_{k j}^{\prime \prime}}  \right)
    \label{eq:net_rate_of_progress}
\end{equation}

\noindent where $\Gamma_j$ is the third-body coefficient, $\mathcal{K}_{fj}$ and $\mathcal{K}_{rj}$ are the forward and reverse reaction coefficients, $\mathcal{C}_k$ is the molar concentration of species $k$, while $\nu_{k j}^{\prime}$ and $\nu_{k j}^{\prime \prime}$ are the forward and reverse stoichiometric coefficients.

Within the viscous sub-layer, it is apparent that reactions involving HCO are dominant. Specifically, the recombination of CO and H to form HCO using H$_2$O as collision partner as well as the recombination of HCO and H to CO and H$_2$ demonstrate the largest rates of progress. The net effect is a net production rate for HCO and a subsequent increase in its mass fraction as shown in Fig. \ref{fig:SpeciesYplusMinor}. At the same time, the recombination of some of the excess HCO with H to form formaldehyde is also activated in this low temperature region. 

The reason for the sharp increase in the mass fractions of HCO and CH$_2$O is the absence of activation energy for the aforementioned reactions. The formation of formaldehyde is driven by a negative activation energy, which promotes the formation of the molecules in low enthalpy environments. As the temperature drops in the vicinity of the wall, the other reaction rates with positive activation energy start decaying, leading to a freezing of the reactions. For the case of HCO and CH$_2$O production reactions however, the rates of progress keep increasing as the density and hence radical concentration increase. This agrees with the results reported by Westbrook et al. \citep{westbrook1981numerical}, who investigated the importance of radical recombination reactions with low activation energy and found that radical recombination reactions become the dominant means of consumption of radical species when the temperatures drop below 700-800 K and the rates of radical-fuel reactions fall to very low values. 

\begin{figure}
\includegraphics[width=.48\textwidth]{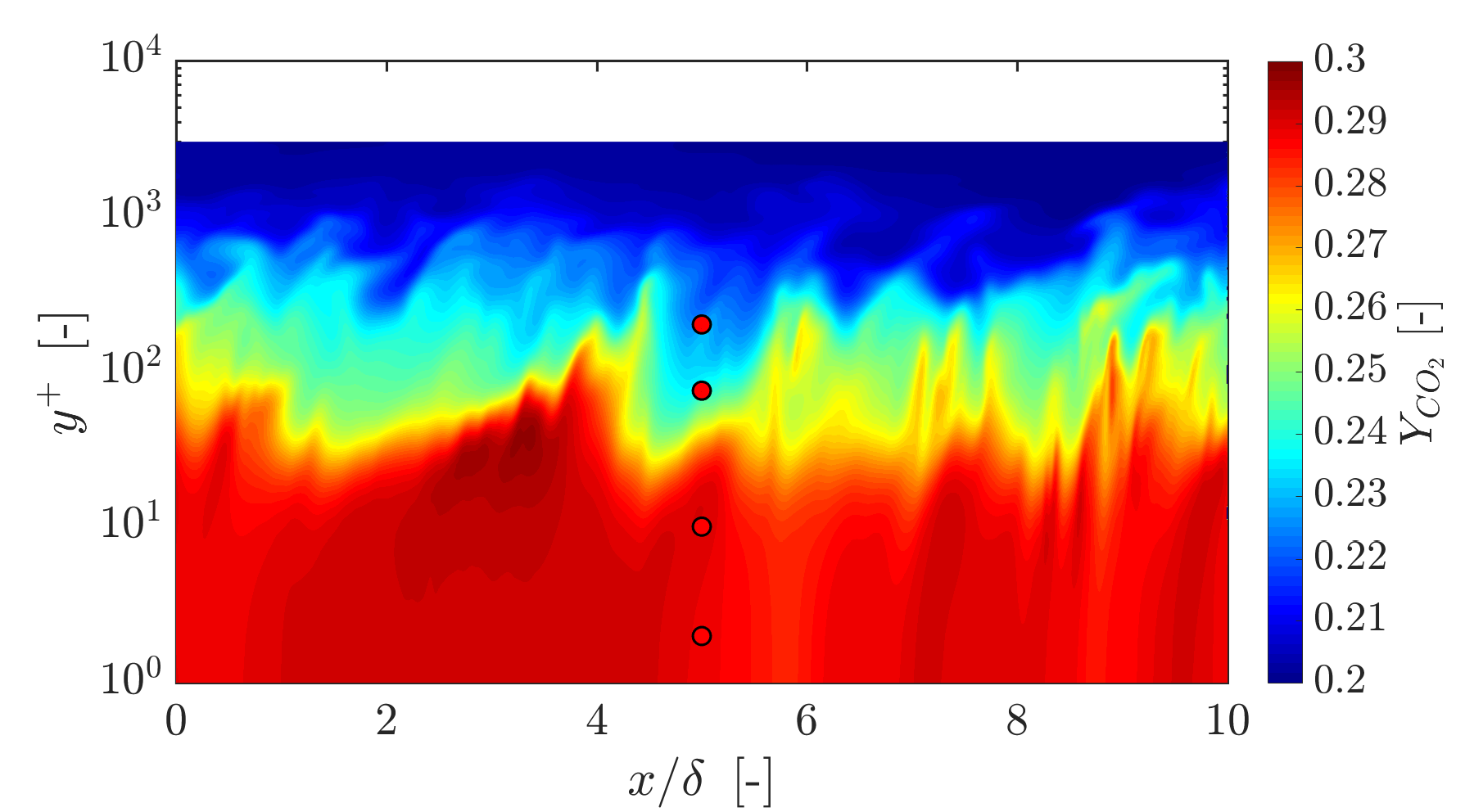}
\caption{\label{fig:CO2_500_2D_log}Instantaneous CO$_2$ species mass fraction for the reacting case with wall temperature equal to 500 K. The wall-normal distances used in Fig. \ref{fig:rxnpath500} are indicated with the red markers. }
\end{figure}
\begin{figure*}
\includegraphics[width=.18\textwidth]{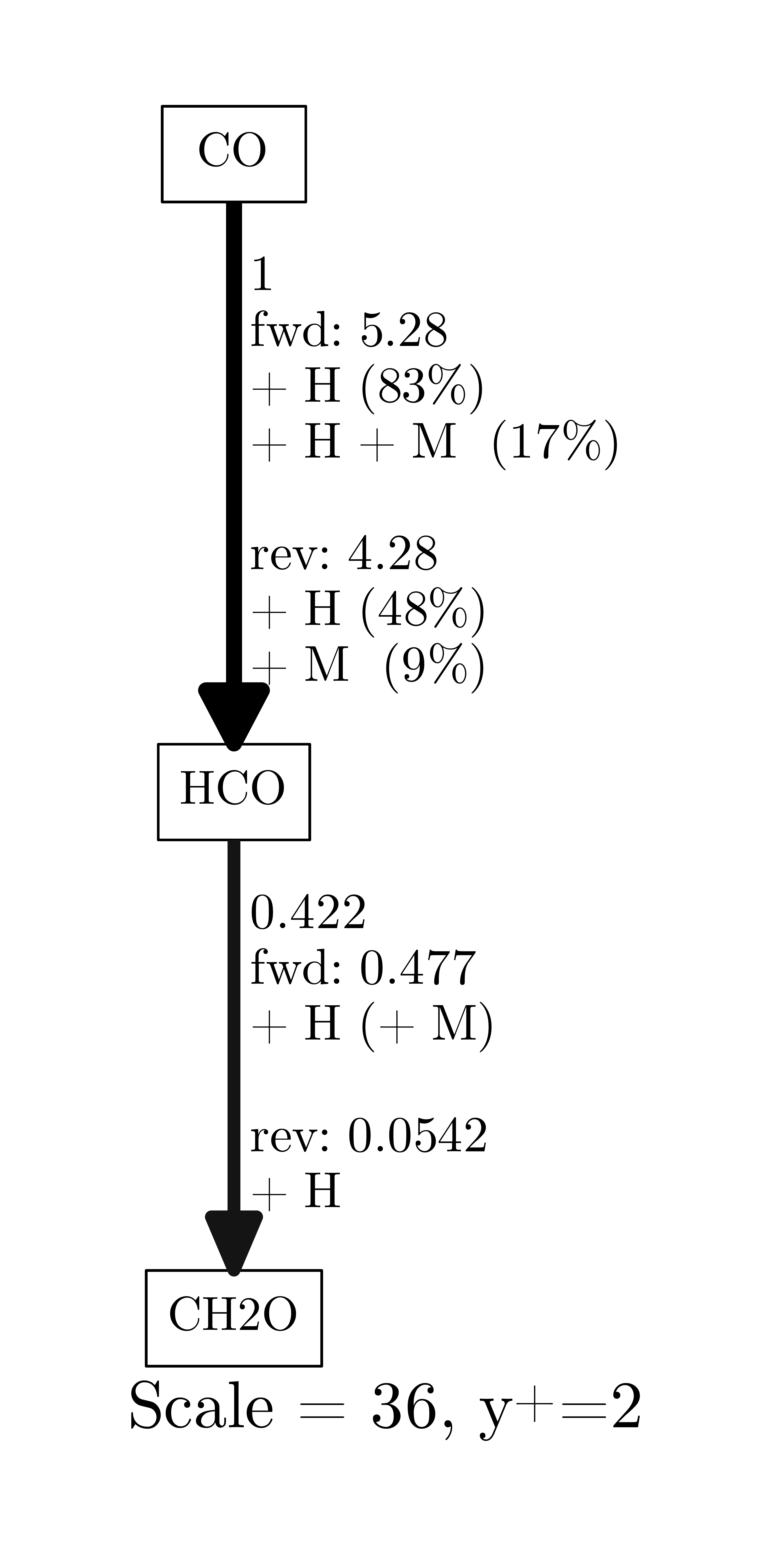}
\includegraphics[width=.24\textwidth]{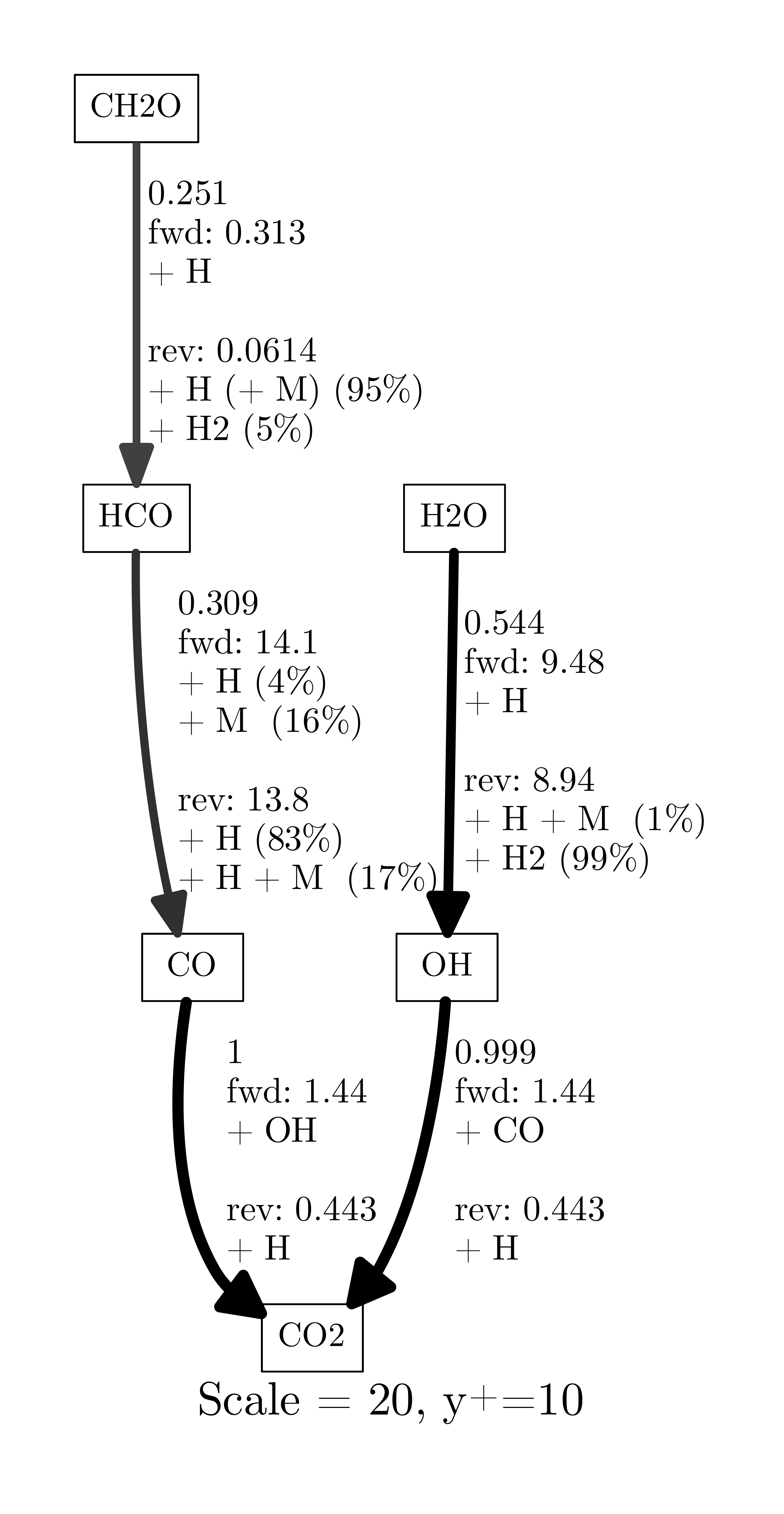}
\includegraphics[width=.18\textwidth]{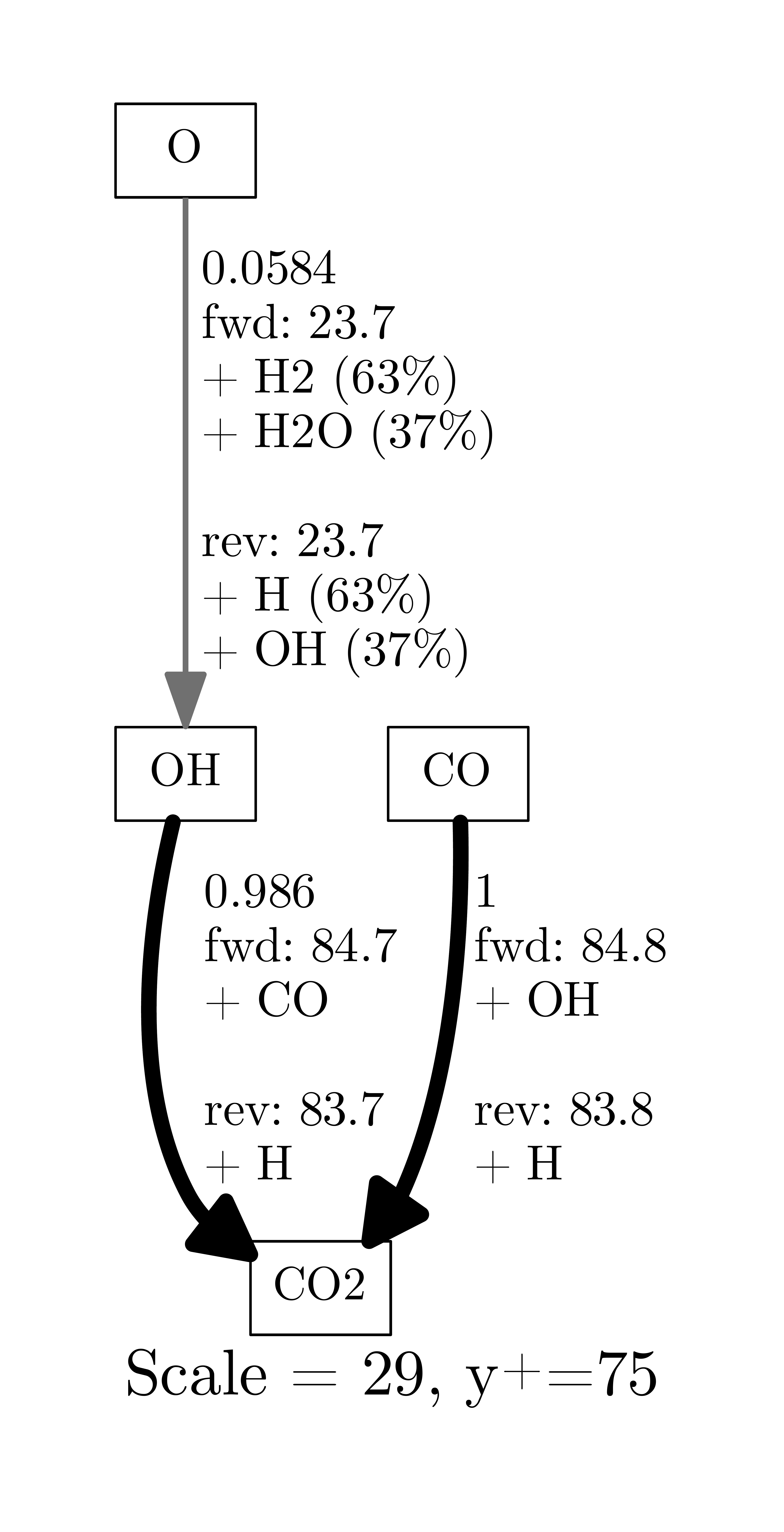}
\includegraphics[width=.38\textwidth]{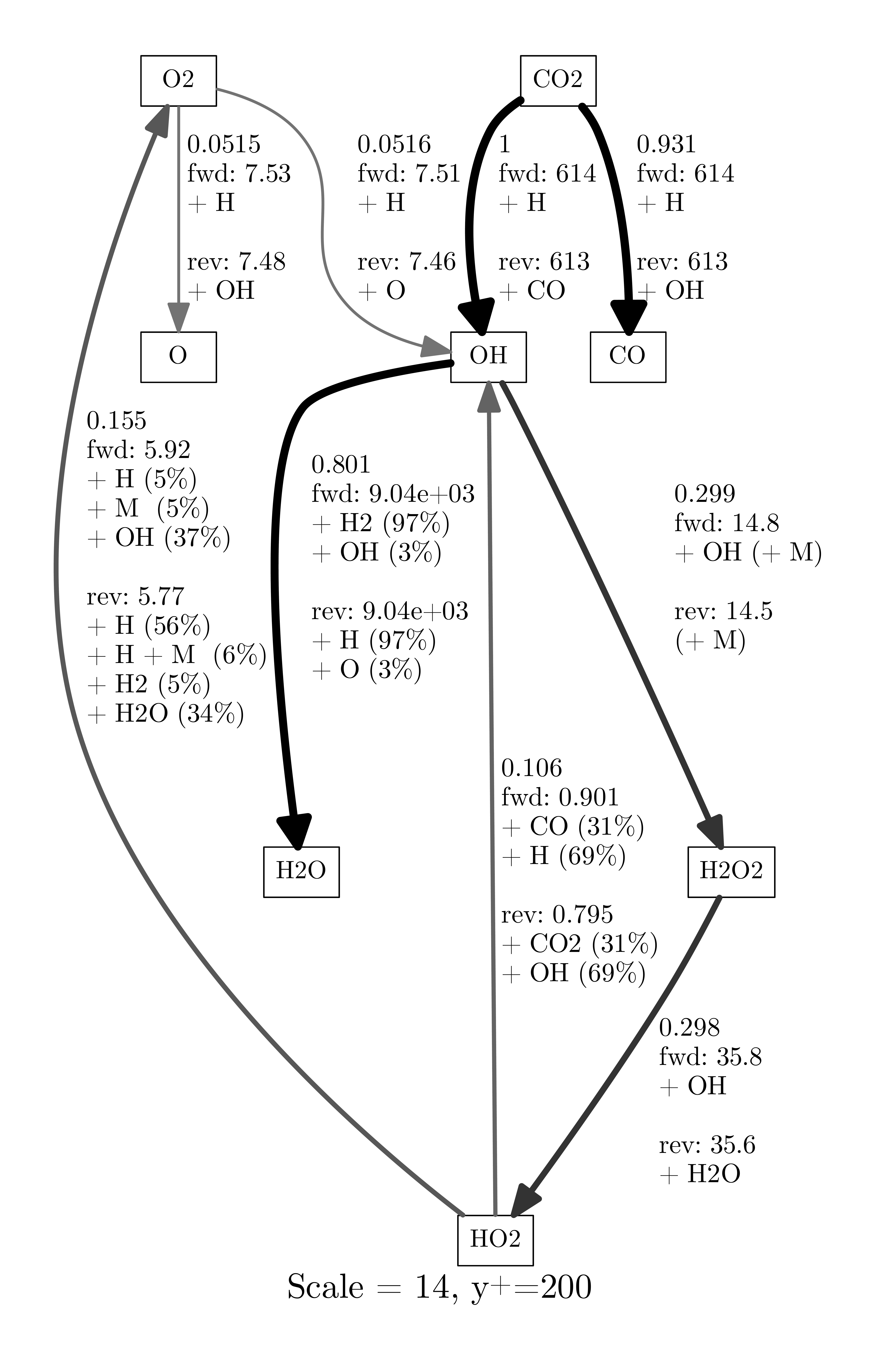}
\caption{\label{fig:rxnpath500} Reaction flux path diagram for the oxygen atom at different wall-distances in the reacting case with $T_w= 500\,$K. }
\end{figure*}

The paths leading to the production of formyl and formaldehyde in the reacting DNS are illustrated in the reaction flux diagram of Figs. \ref{fig:rxnpath500}. Reaction path diagrams at different wall-normal distances for a representative instantaneous snapshot are included. The forward and backward rates of the most dominant reactions are also used as labels in the diagrams. Going from left to right in Fig. \ref{fig:rxnpath500}, the normalized distances $y^+=2$, $y^+=10$, $y^+=75$ and $y^+=200$ are shown. Those distinct locations are also indicated with red markers in Fig. \ref{fig:CO2_500_2D_log}. In the regions further from the wall ($y^+=200$), the production of H$_2$O remains most prominent, as the temperature is still high and the reaction rates of the hydrogen chemistry dominate. Approaching the wall, in the lower temperature environment ($y^+=75$ and $y^+=10$), the recombination of OH and CO to form CO$_2$ is established as the main reaction path. Directly within the viscous sub-layer nevertheless ($y^+=2$), the reaction rates of the CO-to-CO$_2$ recombination die out and CO feeds the production of HCO via the CO + H and CO + H + M paths. Further absorption of a H-atom  leads to the production of formaldehyde. 

\begin{figure}
\includegraphics[width=.4\textwidth]{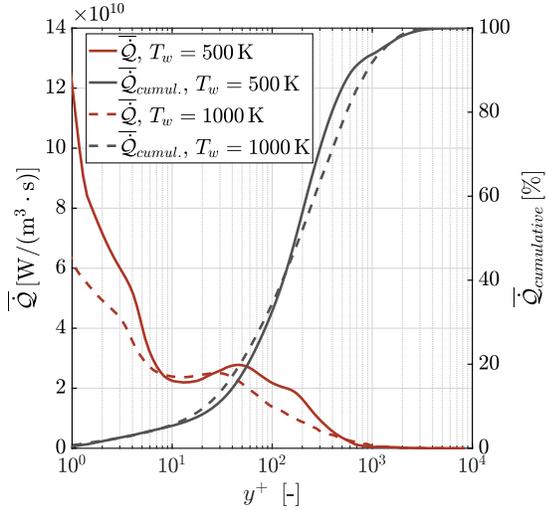}
\caption{\label{fig:Qcumu} Mean heat release rate and cumulative energy release for the reacting cases with $T_w=500\,$K and $T_w=1000\,$K.}
\end{figure}

Given the presence of two distinct reaction zones (one where OH and CO recombine to form carbon dioxide and water and the other where CO recombines to formyl and formaldehyde), the  individual contributions of these two modes to the total heat release rate are quantified in Fig. \ref{fig:Qcumu}. The mean local heat release rate $\overline{\dot{\mathcal{Q}}}$ as well as the integrated cumulative heat release $\overline{\dot{\mathcal{Q}}}_{cumul}$ are shown for the two reacting cases. The cumulative heat release is defined as the percentage of the total energy release starting from the wall and moving towards the free stream:

\begin{equation}
\overline{\dot{\mathcal{Q}}}_{cumul}=\cdot \frac{\int_{0}^{y} \overline{\dot{\mathcal{Q}}}(y) d y}{\int_{0}^{\infty} \overline{\dot{\mathcal{Q}}}(y) d y}
\end{equation}

As expected from the results of the species production rates in Fig. \ref{fig:OmegaDot}, the total heat release rate has a local maximum close to $y^+\approx60$ and a further maximum with a larger value directly at the wall ($y^+=1$). For the case with $T_w=500\,$K, the maximum corresponding to the HCO recombination is more prominent. However, looking at the distribution of $\overline{\dot{\mathcal{Q}}}_{cumul}$ in Fig. \ref{fig:Qcumu}, it becomes evident that the HCO production reactions do not have a significant contribution to the total heat release. Within the viscous sub-layer ($y^+<10$), approximately 6\% of the total energy is set free. The bulk majority of the chemical energy is released between $y^+\approx10$ and $y^+\approx1000$, with a subsequent ceasing of the energy production in the regions with high temperature. The low-temperature recombination reactions with zero activation energy have hence a minimal impact on the total energy release, as they are only activated in a region with restricted volume in the direct vicinity of the wall. 

By comparing the total heat release rate with the net rate of progress for the most dominant reaction in Fig. \ref{fig:OmegaDot}(b), it can be inferred that the two reactions responsible for the majority of the energy being released in the boundary layer are the production of OH following the path H + H$_2$O $\rightarrow$ H$_2$ + OH and the subsequent recombination of OH and CO following the path CO + OH $\rightarrow$ CO$_2$ + H. The latter reaction path has also been identified as the most dominant reaction in the boundary layer of rocket thrust chambers in prior work \citep{perakis2020PCI}.

\subsection{Chemical time-scales}

The exothermic CO recombination reaction that is activated in the low-temperature environment can be explained by Le Chatelier's principle, as the initially equilibrated gas is subjected to a change in enthalpy and reacts by undergoing exothermic reactions that aim at counteracting the applied change. However, with decreasing temperature near the wall, the reaction rates also approach zero which gives rise to two competing effects: the low temperature, which is the initiator for the reactions is also the limiting factor that does not allow the composition to reach its new equilibrium. This results in a significant increase in the chemical time-scales and is expressed by an apparent "quenching" of the chemical reactions.

To understand the competing effects that dominate the change in chemical composition, we choose to describe the chemical time-scale along the wall normal direction using the Computational Singular Perturbation (CSP) technique by Lam et al. \cite{lam1989understanding} which is used in the Chemical Explosive Mode Analysis (CEMA) as introduced by Lu et al. \citep{lu2010three}. The time-scale is then defined as:

\begin{equation}\tau_{c}=\frac{\|\dot{\boldsymbol{\omega}}\|}{\|\boldsymbol{J} \cdot \dot{\boldsymbol{\omega}}\|} \; ,
\label{eq:tau_SPTS}
\end{equation}

\noindent where the species Jacobi matrix is given by the sensitivity of the net species reaction rates to a change in the species mass fractions.

\begin{figure}
\includegraphics[width=.45\textwidth]{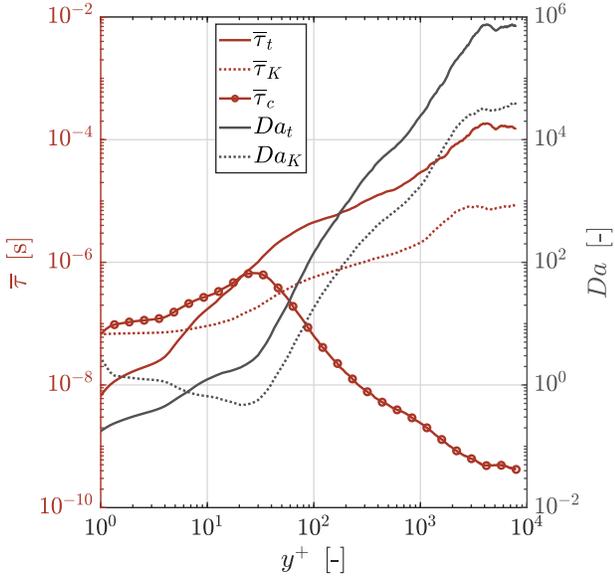}
\caption{\label{fig:tau} Chemical time-scale and Damk\"ohler number for the reacting case with $T_w=1000\,$K. }
\end{figure}

The mean chemical time-scale $\tau_c$ for the reacting cases is shown in Fig. \ref{fig:tau} along with the local Damk\"ohler number. Two different definitions for the Dahmk\"ohler number are included, one based on the Kolmogorov time-scale $\tau_K$ and one based on the integral turbulent time-scale $\tau_t$. The respective time-scales and Damk\"ohler numbers are given as:

\begin{equation}
\tau_K =  \sqrt {\frac{\mu}{\rho  \epsilon} } \; \; , \; \; Da_K =\tau_K / \tau_c
\label{eq:Da_K}
\end{equation}

\begin{equation}
\tau_t =  \frac{k}{\epsilon} \; \; , \; \; Da_t = \tau_t / \tau_c
\label{eq:Da_t}
\end{equation}

Fig. \ref{fig:tau} shows that, the chemical time-scale is small in regions away from the wall, indicating that the gas mixture is close to its equilibrium concentration. Moving towards the wall however, as the temperature decreases, a slowing-down of the reactions takes place. This trend persists up until $y^+\approx 30$, where the chemical time-scale has a local maximum and then subsequently reduces further. This secondary reduction in the time-scale represents the zone where the formyl and formaldehyde production become dominant. At the same time, both the turbulent time-scale and the Kolmogorov time-scale are monotonically reduced for locations closer to the wall. 

This leads to comparable results for $Da_K$ and $Da_t$. Both of them have large values between $4 \times 10^4$ and $9 \times 10^5$ outside the boundary layer and reach values between 0.1 and 5 within the viscous sub-layer. In the region where the reaction rates of the main exothermic recombination reactions of CO and OH start diminishing ($y^+<10$), the Damk\"ohler number reaches values close to unity. This implies that after this point the turbulent and Kolmogorov time-scales dictate the evolution of the species which confirms the assumption that the freezing of the chemical reactions is due to the large chemical time-scales in the low-temperature environment. 

\subsection{Wall heat transfer amplification}

Understanding the conditions that lead to the enhancement of the exothermic recombination reactions in the cooled boundary layer is of major importance for the prediction of wall heat loads in rocket thrust chambers and other propulsion systems. Using the comparison between the inert and reacting cases, the effect of the additional energy release in the wall vicinity can be isolated and quantified. The results for the mean wall heat loads for the different DNS cases examined here are summarized in Table \ref{tab:wall_loads}.

\begin{table}
\caption{\label{tab:wall_loads} Mean wall heat flux for the different cases.}
\begin{ruledtabular}
\begin{tabular}{ll}
\textbf{Case} & \textbf{Mean heat flux} \\
 Reacting, $T_w=500\,$K & 11.93 $\mathrm{MW/m^2}$\\
 Reacting, $T_w=1000\,$K & 9.87 $\mathrm{MW/m^2}$\\
 Inert, $T_w=500\,$K & 10.01 $\mathrm{MW/m^2}$\\
 Inert, $T_w=1000\,$K & 8.27  $\mathrm{MW/m^2}$\\
\end{tabular}
\end{ruledtabular}
\end{table}

These results show that, lower wall temperatures lead to a higher heat flux, as this is proportional to the temperature difference between the wall and the free stream temperature (which is equal to the adiabatic equilibrium temperature). The same is true when comparing the two reacting cases to each other, with the lower wall temperature leading to a higher wall heat transfer rate. For a given wall temperature, an increase of approximately 20\% is obtained from the effect of the chemical recombinations. This degree of under-prediction of the heat loads when using models that do not account for recombination reactions is comparable with the values reported by Betti et al. \citep{betti2016chemical} for low-pressure methane/oxygen rocket engines. It was also reported in the same work that the difference between inert and reacting treatment for the boundary layer reduces with increasing operating pressures.

\begin{figure}
\includegraphics[width=.48\textwidth]{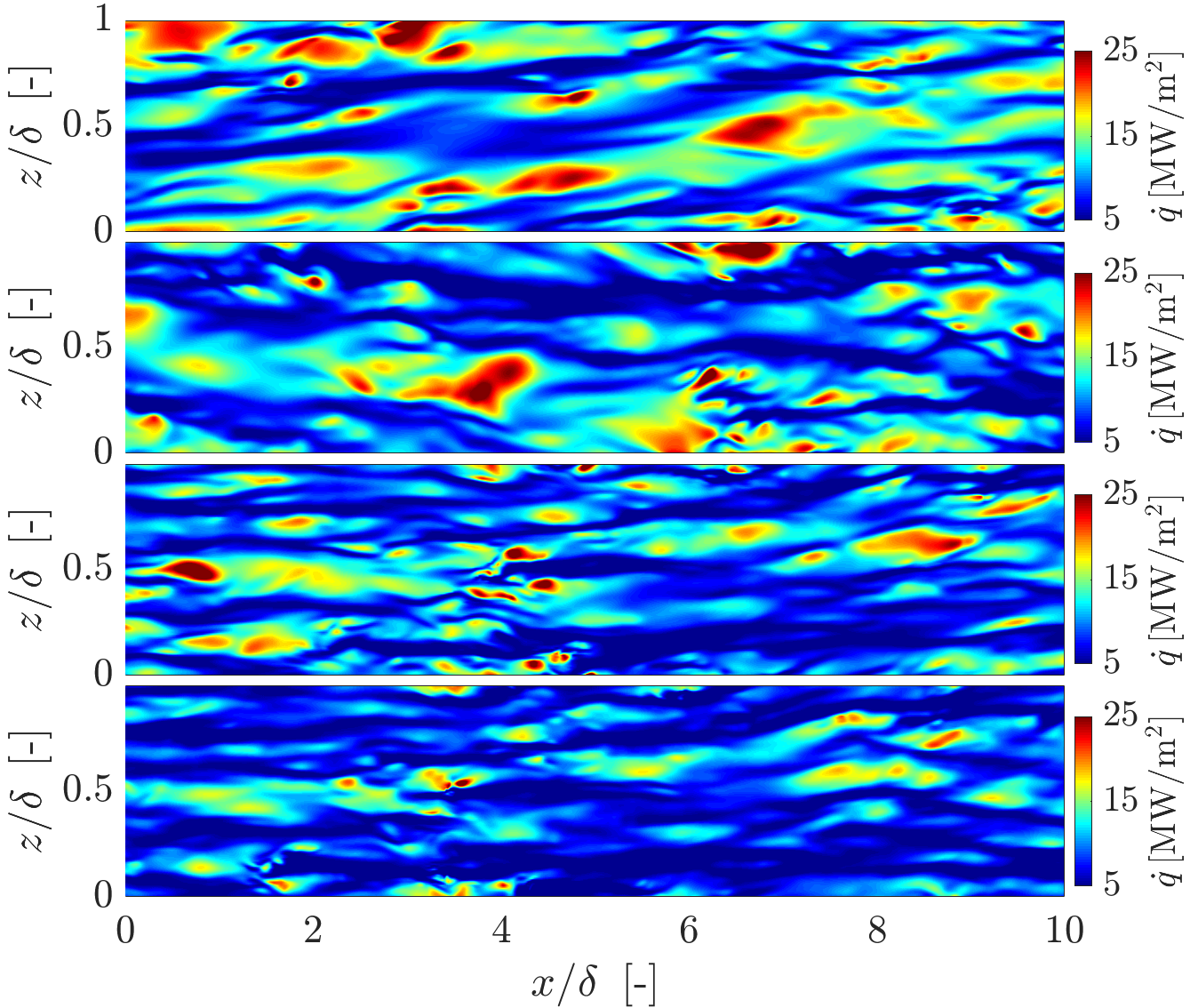}
\caption{\label{fig:wallheatflux} Instantaneous wall heat flux. From top to bottom: reacting case with $T_w=500\,$K, reacting case with $T_w=1000\,$K, inert case with $T_w=500\,$K, inert case with $T_w=1000\,$K}
\end{figure}

\begin{figure}
\includegraphics[width=.4\textwidth]{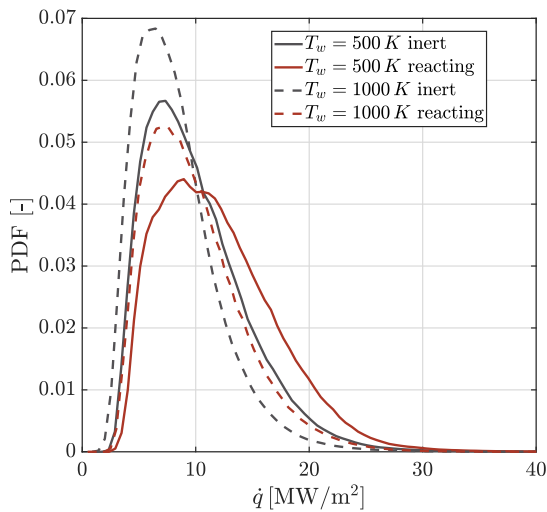}
\caption{\label{fig:PDF}PDFs for the wall heat flux distribution.}
\end{figure}

\begin{figure*}
\includegraphics[width=.85\textwidth]{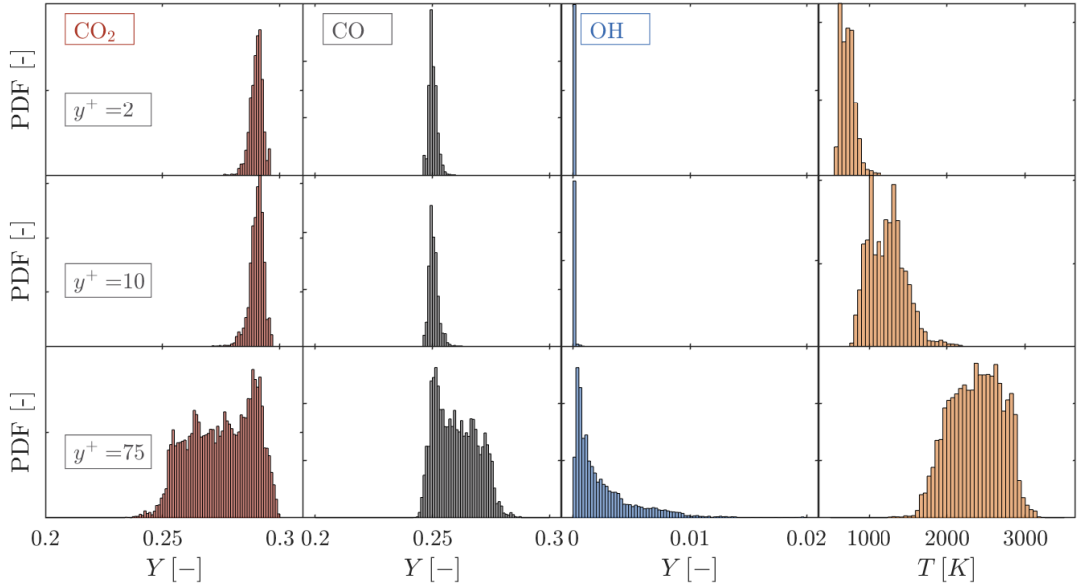}
\caption{\label{fig:PDF_Y} PDFs of species mass fractions and temperature at different wall distances for the reacting case with $T_w=500\,$ K. }
\end{figure*}

Apart from the averaged values for the heat loads reported in Table \ref{tab:wall_loads}, the local variation of the heat flux is plotted in Fig. \ref{fig:wallheatflux} for representative instantaneous snapshots. Localized "hot islands" with heat flux values two times larger than the mean average are visible. This is also illustrated in Fig. \ref{fig:PDF}, where the Probability Density Function (PDF) of the instantaneous local heat flux values is plotted for each of the four simulations. A consequence of the overall higher heat flux level, the reacting cases have a wider distribution compared to the respective inert ones. The presence of strong fluctuations in the experienced wall heat loads can lead to significant deviations from the mean values which in turn could lead to strong temperature gradients and higher thermal stresses on the walls. However, the time-scales of the turbulent fluctuations which are responsible for the spatial variation, shown in Fig. \ref{fig:wallheatflux} and \ref{fig:PDF}, are much shorter than typical conduction time-scales in the material and hence do not pose an issue for the design of cooling systems.

\subsection{Turbulent fluctuations}

\begin{figure*}
\includegraphics[width=.85\textwidth]{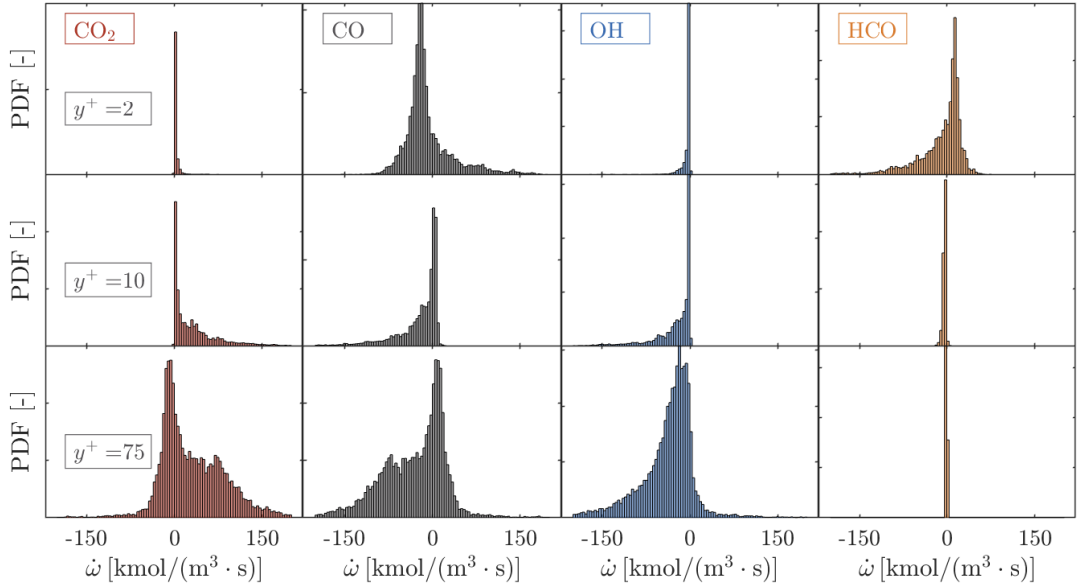}
\caption{\label{fig:PDF_omega} Probability density function of species molar reaction rates at different wall distances for the reacting case with $T_w=500\,$ K. }
\end{figure*}

In an effort to examine the sensitivity of reaction rates to compositional variations, Figs. \ref{fig:PDF_Y} and \ref{fig:PDF_omega} show PDFs of $Y_k$, $T$ and $\dot{\omega}_k$ at three distinct distances from the wall. The case with $T_w=500\,$K is shown here and the results obtained with the higher wall temperature are qualitatively similar. In contrast to the enthalpy-conditioned distribution from Fig. \ref{fig:Scatter}, which clearly showed a minimal variation of the species mass fractions around its average, the PDFs at different spatial locations show a much larger variability. Wider distributions for both the temperature as well as major species mass fractions are observed within the logarithmic region of the boundary layer ($y^+=75$). Approaching the wall and entering the viscous sub-layer where turbulent effects are less prominent, the width of the distributions quickly shrinks. For OH in fact, the PDF of the mass fraction converges to a Dirac function around zero, as most of the OH is consumed before the begin of the viscous sub-layer.

The net molar production rates at $y^+=75$ demonstrate a clear correlation between the consumption of CO and the production of CO$_2$ and exhibit a large variation which extends in the range from $-$150 to 150 $\mathrm{kmol/(m^3 s)}$, while the average value lies at $\pm$ 35 $\mathrm{kmol/(m^3 s)}$ for CO$_2$ and CO, respectively. Closer to the wall, the correlation between the CO and CO$_2$ reaction rates ceases to be dominant, as carbon dioxide no longer undergoes any further chemical reactions in alignment with the results in Fig. \ref{fig:OmegaDot}. Instead, as the reaction path including HCO is activated, a correlation between the reaction rates of formyl and carbon monoxide is visible. It is also noteworthy that the reaction rate of HCO was close to zero for locations within the log-layer, whereas it exhibits a large variability in the vicinity of the wall owing to non-negligible variations in temperature and CO mass fraction.

\begin{figure*}
\includegraphics[width=.45\textwidth]{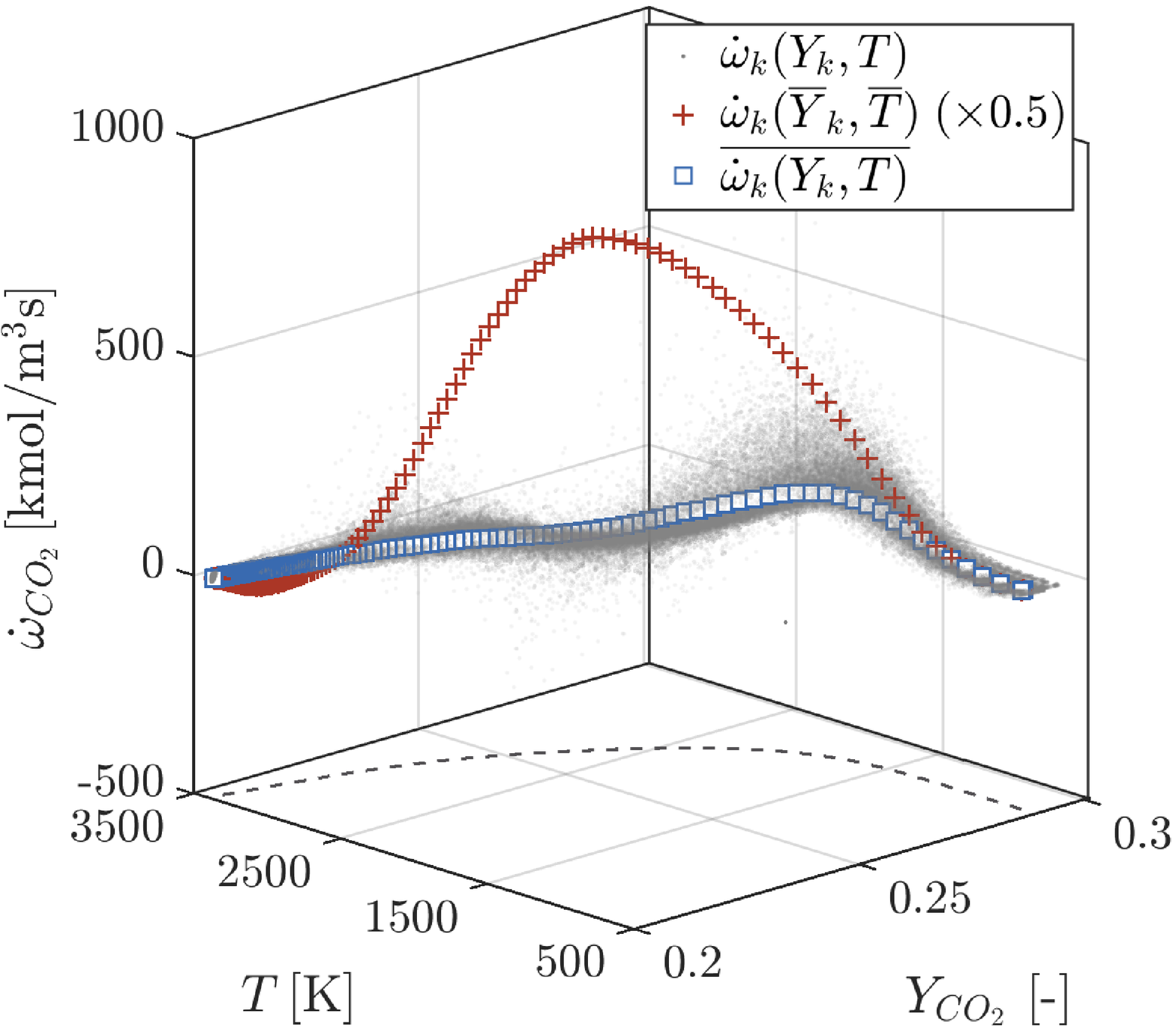}
\includegraphics[width=.45\textwidth]{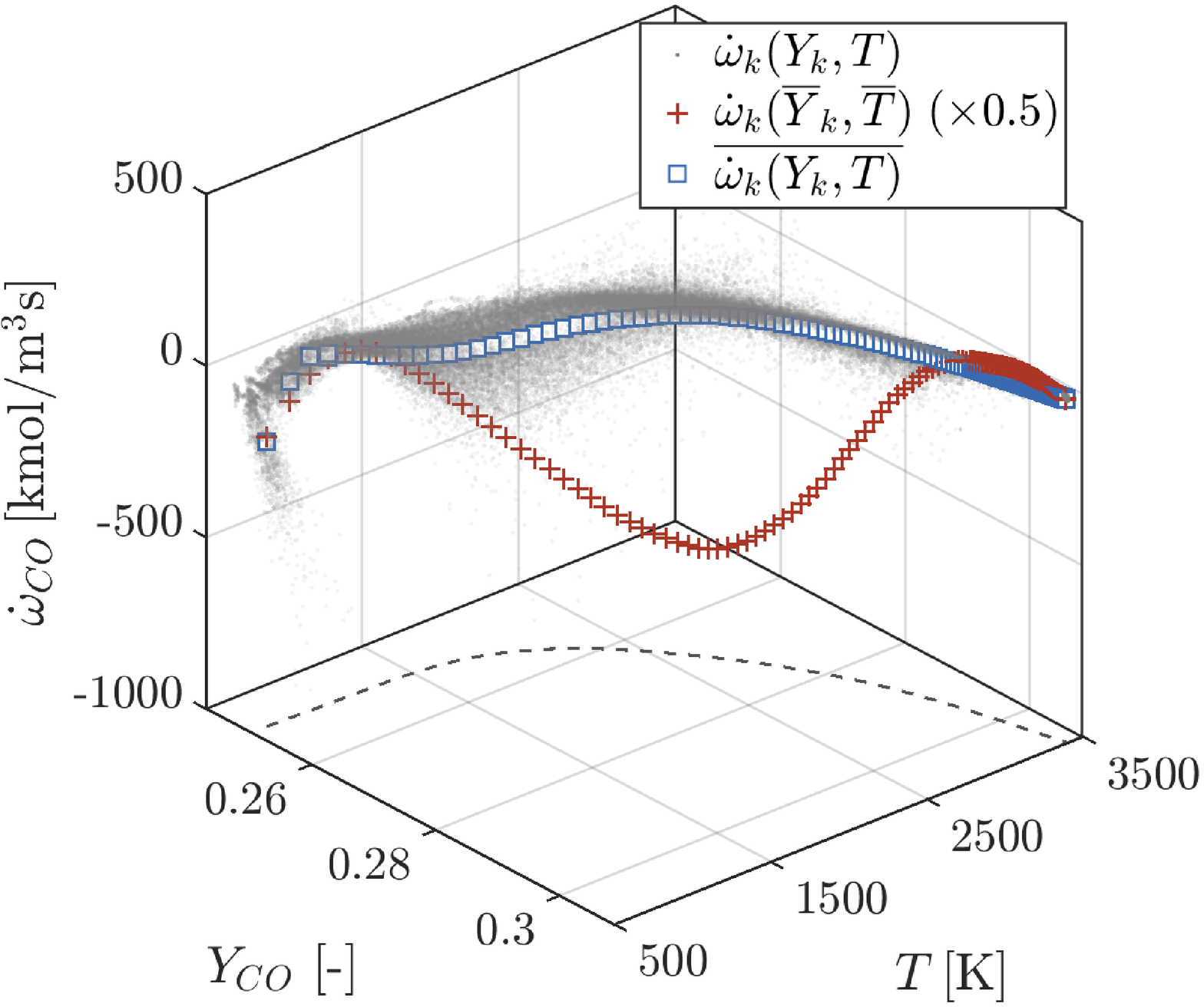}
\caption{\label{fig:PR_conditional} Scatter plot of the instantaneous and averaged chemical source terms for (a) CO$_2$ and (b) CO in conditional space for $T_w= 500\,$K. The dashed lines correspond to the manifold correlating mean temperature and mean mass fractions. }
\end{figure*}

To further examine the non-linearity of mass fractions and reaction rates, a scatter plot of instantaneous and averaged net species production rates are given in Fig. \ref{fig:PR_conditional}. The grey points represent instantaneous results, mapped onto the conditional $Y_k-T$ space, while the dashed line corresponds to the projection of the averaged quantities, i.e. to the manifold $\overline{Y}_k - \overline{T}$. 

It is evident that the estimation of the reactions rates using the averaged fields for mass fractions and temperature $\dot{\omega}_k(\overline{Y}_i,\overline{T})$ leads to significant overestimation compared to the time-averaged production rates $\overline{\dot{\omega}_k(Y_i,T)}$ (which correspond to the results from Fig. \ref{fig:OmegaDot}) both for CO and CO$_2$. The two results for the production rates deliver similar values in the hot regions (where the gas is near its chemical equilibrium) as well as in the direct vicinity of the wall, where a quenching of the reactions occur. For the intermediate positions however, differences larger than one order of magnitude are observed. This is an indication for strong turbulence-chemistry interaction, which leads to the confirmation that the closure of the chemical source term is inadequately described by a first-order approximation of the reaction rate:

\begin{equation}
\overline{\dot{\omega}_k(Y_i,T)} \neq \dot{\omega}_k(\overline{Y}_i,\overline{T})
\end{equation}

\begin{figure}
\includegraphics[width=.4\textwidth]{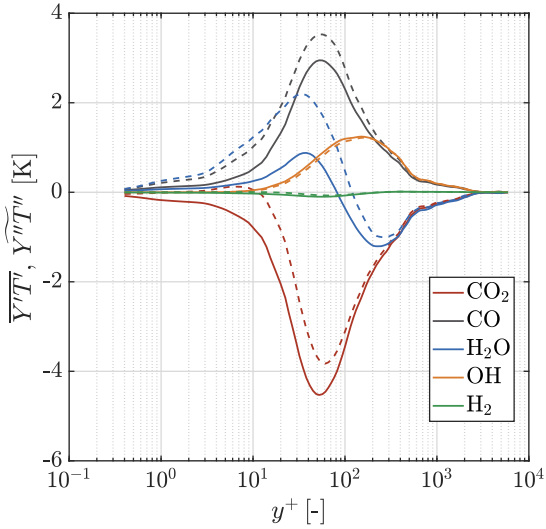}
\caption{\label{fig:YprimeTprime} Correlation of turbulent fluctuations for species mass fractions and temperature for the reacting case with $T_w=500\,$K. The solid line represents the Reynolds averaged quantity $\overline{Y^{\prime}T^{\prime}}$, while the dashed line represents the Favre averaged $\widetilde{Y^{\prime \prime}T^{\prime \prime}}$. }
\end{figure}

The deviation between the two quantities can be attributed to the correlation between mass fraction fluctuations and temperature fluctuations. This is shown in Fig. \ref{fig:YprimeTprime}, which clearly illustrates that a positive correlation holds in the reacting zone for CO, while a negative correlation can be found for CO$_2$. No significant difference can be found between the compressible and incompressible definitions ($\widetilde{Y^{\prime \prime}T^{\prime \prime}}$ and $\overline{Y^{\prime}T^{\prime}}$), as they both give qualitatively identical results.

Finally, we investigate whether the turbulent fluctuations and the feedback produced by the chemical reactions give rise to an enhancement of the turbulent heat transfer. To quantify this, the turbulent Prandtl number is calculated for all four DNS cases. The turbulent Prandtl number is defined as the ratio of the turbulent thermal diffusivity $\alpha_T$ and the turbulent viscosity $\alpha_M$. The strong Reynolds analogy assumes that the turbulent heat transfer and the turbulent momentum transfer are similar, resulting in a turbulent Prandtl number equal to unity. Based on experimental and numerical studies, it has been found that the Reynolds analogy is valid for most boundary layer flows, although departures from $Pr_t=1.0$ have been reported \citep{kays1994turbulent}.

Common practice in most turbulence models for RANS, is to approximate $Pr_t$ by a constant value, which in some cases, can lead to incorrect heat transfer predictions. This is typically the case in supercritical flows, where strong gradients in the molecular Prandtl number give rise to spatial variations for $Pr_t$ \citep{yoo2013turbulent}. For gases at high-temperature conditions, with nearly constant molecular Prandtl number (as in the present operating point) and high turbulent Peclet number, the Prandtl number is nearly constant within the log region with values around 0.85 and 1, while in the "wake" region of external turbulent boundary layers it was found to decrease in the neighborhood of 0.5-0.7 \citep{kays1994turbulent}. Previous studies of reacting turbulent boundary layers performed by Cabrit et al. \citep{cabrit2009direct} found that the turbulent Prandtl number varies from around 0.5 in the middle of their turbulent channel up to values close to 1.2 in the wall vicinity. No qualitative differences were obtained in the $Pr_t$ profiles between inert and reacting cases.

\begin{figure}
\includegraphics[width=.4\textwidth]{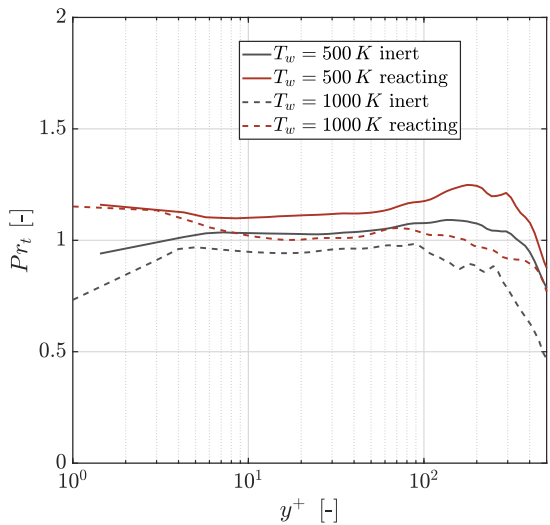}
\caption{\label{fig:Prt} Turbulent Prandtl number for the four DNS cases as a function of the wall-normal distance. }
\end{figure}

Similar findings are reported in Fig. \ref{fig:Prt} by analyzing the results of the present DNS. The compressible definition was employed for the calculation:

\begin{equation}P r_{t}=\frac{\widetilde{u^{\prime \prime} v^{\prime \prime}}}{\widetilde{v^{\prime \prime} h^{\prime \prime}}} \frac{\partial \widetilde{h} / \partial y}{\partial \widetilde{u} / \partial y}
\label{eq:Prt}
\end{equation}

For all cases, the turbulent Prandtl number appears to be nearly constant throughout the logarithmic and viscous sub-layer regions. Differences between the individual cases are small, with the reacting cases having the tendency of producing higher turbulent Prandtl numbers. Values between 0.9 and 1.2 are found within the region where the bulk energy release takes place ($y^+\approx 10$ to $y^+\approx 100$). Further from the wall, a drop of the turbulent Prandtl number to values as low as 0.5 is observed. 

\section{Conclusions}

In the present study, DNS of a reacting boundary layer have been carried out. Mixtures representative of the conditions within rocket engines and high-pressure propulsion systems have been chosen, corresponding to an equilibrated mixture of methane and oxygen at 20 bar. In order to isolate effects of chemical reactions resulting from the low-enthalpy environment in the vicinity of the isothermal wall, inert and reacting simulations were carried out for the same operating conditions. A detailed chemical mechanism was employed to account for the role of chemical radical recombinations in the vicinity of the wall. 

Evaluation of the temperature profile showed that deviations between the inert and reacting cases were prominent, resulting from the energy release within the boundary layer. The energy released from the recombination of CO and OH to form H$_2$O and CO$_2$ was identified as the main effect for the temperature increase. This effect was found to be qualitatively similar for both wall conditions.

The analysis of reaction rates showed strong non-equilibrium effects. Although a clear correlation between the degree of recombination of CO and the local enthalpy value was found, the species mass fractions profiles deviated from the theoretical equilibrium values. Moreover, for  regions up until the begin of the viscous sub-layer ($y^+\approx 10$), a quenching of the major species mass fractions was observed. Further examination of the chemical reaction rates confirmed that the occurrence of this layer with chemically quenched composition arises from the low temperature and resulting long chemical time-scales. In fact, the chemical time-scales at the locations where the recombination rates start diminishing reach values that are larger than the turbulent and Komlmogorov time-scales and hence chemistry is no longer the rate-defining process.


Closer to the wall, despite the termination of the CO recombination, the recombination of CO to HCO and the subsequent formation of CH$_2$O is activated. This reaction path was detected using reaction flux diagrams and showed that the production of formyl and formaldehyde is favored by low wall temperatures. Due to the restricted extent of the region in which the temperatures are low enough to activate the aforementioned reactions, the total energy released from this reaction path amounted to approximately 5\% of the total heat release in the boundary layer. 

As far as the wall heat transfer is concerned, the exothermic reactions contributed to an additional 20\% in heat flux compared to the equivalent inert cases. No qualitative difference was observed between the results with low (500 K) and high (1000 K) wall temperature. The non-negligible augmentation of the heat transfer puts additional emphasis on the importance of the recombination reactions in modeling efforts of flame-wall interaction, like in the case of non-adiabatic flamelet manifolds. 

A strong coupling between turbulence and chemistry was also inferred from the results. Strong variations in mass fractions and temperature within the reacting zone resulted in a large variance of the resulting reaction rates and broad PDFs for the source terms. The assumption of first-order representation of the reaction rate was hence assessed as inadequate to describe the turbulence-chemistry interaction processes in the reaction zone. 
Despite the strong correlation of species and temperature fluctuations and the additional energy released via the chemical reactions, no significant enhancement of the turbulent heat transfer was observed. Evaluation of the turbulent number for both the reacting and inert cases delivered values for $Pr_t$ close to unity with fairly constant profiles along the entirety of the logarithmic and viscous sub-layers.


The strong presence of non-equilibrium effects, dictated by the competition between turbulent and chemical time-scales as well as the intensity of the TCI throughout the turbulent boundary layer, highlight the difficulty of modeling the near-wall region in the presence of chemically reacting flows and strong temperature gradients. Examination of further operating points and fuels is of interest in order to understand the processes occurring within the reacting boundary layer of rocket engines and gas turbines for a broad spectrum of applications. Based on the representative load point chosen for this analysis however, the importance of the recombination effects on the wall heat flux augmentation is illustrated. For that reason, future models developed for the treatment of near-wall effects in reacting simulations should account for the quenching of the reactions within the buffer and viscous sub-layers.

\section{Acknowledgements}
\begin{acknowledgments}
Resources supporting the DNS study were provided by
the NASA High-End Computing Program through the NASA Advanced Supercomputing Division at Ames Research Center (Award No. NNX15AV04A) and by the GCS Supercomputer SuperMUC at Leibniz Supercomputing Centre. Financial support has
been provided by the German Research Foundation (Deutsche Forschungsgemeinschaft DFG) in the framework of the Sonderforschungsbereich Transregio
40.
\end{acknowledgments}


\bibliography{DNS_bibliography}

\end{document}